\documentclass[eqsecnum,preprint,preprintnumbers,12pt,aps]{revtex4}
\usepackage{subeqnarray}

\begin{document}

\title{Renormalized  Effective Actions in Radially Symmetric
Backgrounds I: Partial Wave Cutoff Method}

\author{Gerald V. Dunne}
\email{dunne@phys.uconn.edu}
\affiliation{
Department of Physics, University of Connecticut,
Storrs, CT 06269, USA;\\
and Centre for the Subatomic Structure of Matter, Department of Physics, University of Adelaide, SA 5005, Australia}
\author{Jin Hur} 
\email{hurjin2@snu.ac.kr}
\author{Choonkyu Lee} 
\email{cklee@phya.snu.ac.kr}
\affiliation{Department of Physics and Center for Theoretical
Physics\\ Seoul National University, Seoul 151-742, Korea}

\begin{abstract}
The computation of the one-loop effective action in a radially symmetric background can be reduced to a sum over partial-wave contributions, each of which is the logarithm of an appropriate one-dimensional radial determinant. While these individual radial determinants can be evaluated simply and efficiently using the Gel'fand-Yaglom method, the sum over all partial-wave contributions diverges. A
renormalization procedure is needed to unambiguously define the finite renormalized effective action.
Here we use a combination of the Schwinger proper-time method, and a resummed uniform DeWitt expansion. This provides a more elegant technique for extracting the large partial-wave contribution, compared to the higher order radial WKB approach which had been used in previous work. We illustrate the general method with a complete analysis of  the  scalar one-loop effective action in a class of radially separable SU(2) Yang-Mills background fields.
We also show that this method can be applied  to the case where the background gauge fields have asymptotic limits appropriate to uniform field strengths, such as for example in the Minkowski solution, which describes an instanton immersed in a constant background.
Detailed numerical results will be presented in a sequel.

\end{abstract}
\maketitle

\section{Introduction}

In the study of quantum field theories, one is often led to
consider the one-loop effective action in nontrivial background
fields. While the renormalization counterterm structure of
one-loop effective actions can readily be exhibited for general
backgrounds, the explicit evaluation of its full finite part in an
interesting specific background still constitutes a highly
nontrivial problem. For gauge theories, explicit analytic results are
known only for very special backgrounds:  the Euler-Heisenberg effective
action for a background with constant abelian field strength 
\cite{heisenberg, schwinger, dunne},  its generalization to a 
nonabelian covariantly constant background 
\cite{duff,batalin,yildiz,leutwyler}, and a special solvable 
abelian background
\cite{nikishov,devega,cangemi,hall}.
For applications in both continuum and lattice field theory, one would 
like to enlarge this set of backgrounds for which we have accurate 
computations of the finite renormalized effective action.

In a series of recent publications (together with Hyunsoo Min)
\cite{idet}, we presented a new method for computing the renormalized 
one-loop effective
action in a radially symmetric nonabelian background, and used it to 
evaluate
explicitly the QCD single-instanton determinant for arbitrary quark mass
values. The related computation in the massless limit was
performed in a classic paper of 't Hooft \cite{thooft}, 
while the heavy quark mass limit was studied in
\cite{novikov,kwon}. The new method in \cite{idet} works for any 
quark mass, not relying on small or large mass expansions, and the 
result interpolates smoothly and precisely between these two extremes. 
In this paper we present the general formalism, and we introduce a 
simplified analysis based on a uniform Schwinger-DeWitt expansion, 
replacing the higher order radial WKB analysis used in \cite{idet,wkbpaper}.
This approach has also been used to evaluate the exact determinant 
prefactor in false vacuum decay (or nucleation) \cite{fvdecay}. We also 
note that a related method has recently been applied to the 
two-dimensional chiral Higgs model \cite{burnier}. Finally, an 
alternative derivation, using zeta functions rather than a partial-wave 
cutoff, of the determinant of a radially symmetric  Schr\"odinger operator 
has been given in \cite{dk}.

The starting idea is very simple. If the background field is radially symmetric, the
effective action $\Gamma$ can be expressed formally in terms
of one-dimensional functional determinants of radial differential
operators for various partial waves. Explicitly, writing $J$ for
all quantum numbers specific to a given partial wave, it has the
general structure
\begin{equation}
\label{generalgammastructure}
\Gamma \sim \sum_{J=0}^\infty \ln\left(\frac{ \det ({\cal H}_J + m^2 )}
{ \det ({\cal H}^{\rm free}_J + m^2 )}\right)\quad .
\end{equation}
Here ${\cal H}_J$, the radial Schr\"{o}dinger-type differential
operator for the $J$-th partial wave, contains a nontrivial
(background-dependent) potential term, while ${\cal H}^{\rm free}_J$ 
is the corresponding free operator. In general there will also be  appropriate degeneracy 
factors in the sum.
For a given partial wave $J$, the individual determinant is finite once we divide
by the corresponding `free' contribution. These finite one-dimensional determinants can 
be evaluated easily using the Gel'fand-Yaglom method 
\cite{gy,levit,coleman,dreyfus,forman,kirsten,kleinert}, which reduces the computation to a 
trivial (numerical) integration with initial value boundary conditions.

The non-trivial aspect of this approach is that the sum over partial waves $J$ diverges 
(in spacetime dimension $d\geq 2$). This is of course because the formal sum in 
(\ref{generalgammastructure}) neglects renormalization. We solve this problem by 
introducing a partial-wave cutoff $J_L$, and we isolate the divergence of the sum in 
a form that can be absorbed via renormalization. Specific details are presented in the 
body of this paper, but the general structure is that we write the sum as
\begin{eqnarray} \label{renormalizationform}
\Gamma&=&
 \sum_{J=0}^{J_L} \ln\left(\frac{ \det ({\cal H}_J + m^2 )}{ \det ({\cal H}^{\rm free}_J + m^2 )}\right)
 + \sum_{J>J_L}^\infty \ln\left(\frac{ \det ({\cal H}_J + m^2 )}{ \det ({\cal H}^{\rm free}_J + m^2 )}\right)
+ \textrm{(counterterm)} \quad.
\end{eqnarray}
The finite, renormalized effective action is then evaluated as follows:
\begin{itemize}
\item
The first sum is evaluated numerically using the Gel-fand Yaglom method.
\item
The second sum is evaluated analytically in the large $J_L$ limit, after regulating the determinants. This step uses our uniform Schwinger-DeWitt expansion and Euler-Maclaurin summation.
\item
The analytic large $J_L$ behavior of the regulated determinants leads to the correct renormalization counterterm, and moreover  cancels exactly the numerical divergences of the first  sum as $J_L\to\infty$. This produces a finite renormalized answer.
\end{itemize}
The technically difficult part of the computation is the  analytic computation of the large $J_L$ behavior of the second sum in (\ref{renormalizationform}). This was achieved in \cite{idet} using second-order radial WKB, based on Dunham's formula \cite{dunham}. While this is very general, it can be quite cumbersome for complicated background fields. In this paper we present a simpler method to implement this part of the computation. The
analysis reduces to simple algebraic manipulations, and can be more readily generalized. With this new approach we can now evaluate exactly the finite renormalized effective action for a  very general class of
radially-symmetric backgrounds. In fact, with a background field
involving an {\it unspecified} radial function, the large
partial-wave contribution to the renormalized effective action can
now be evaluated explicitly. This will be important to discuss the
background-field-dependence of the effective action and also to test
various approximation schemes.

This paper is organized as follows. In Section
\ref{sectionradial} we define the renormalized one-loop effective
action for the case of a scalar field in a class of
spherically-symmetric Yang-Mills background fields, assuming four dimensional
Euclidean spacetime. Its Schwinger proper-time representation
\cite{schwinger} is given. The full amplitude is then expressed
using partial wave amplitudes, and we also elaborate here on the
role of the two kinds of proper-time Green functions, one related to
the quadratic differential operator given in 4-D spacetime and the
other for the radial quadratic differential operator. In Section
\ref{sectionlarge} we explain how the large-$l$ partial wave
contribution to the full effective action can be evaluated
explicitly using a generalized DeWitt WKB expansion for the radial
proper-time Green function (i.e., the $\frac{1}{l}$-expansion) and
the Euler-Maclaurin summation method. This allows us to present the
renormalized one-loop effective action (in the class of spherically
symmetric backgrounds) in a form amenable to direct numerical
analysis. In Section \ref{constantsection} we show that our
$\frac{1}{l}$-expansion formula can be applied to the calculation of
the large partial-wave contribution even when background gauge
fields do not fall off at large distance but approach those of
uniform field strength. Also given here is the exact
partial-wave-based treatment of the effective action in the
background corresponding to strictly uniform self-dual field
strengths. In Section \ref{discussion} we conclude with some
relevant discussions and comments.
There are several appendices which contain supplementary materials
and some technical details.
In Appendix
\ref{appfree} the explicit form of the free radial proper-time Green
function in $n$ spacetime dimension is considered. In Appendix
\ref{appmatrixdewitt} we study the coefficient functions in the
$\frac{1}{l}$-expansion when the potential is matrix-valued.
Appendix \ref{appeuler} contains a brief account of the
Euler-Maclaurin summation formula, and
some explicit results obtained using this formula in
connection with our problem.
In a sequel we will address matrix-valued problems in more detail, and 
present detailed numerical results for the general radial cases for which the 
formalism is developed in this current paper.

\section{Effective Action in Radially Symmetric Backgrounds}
\label{sectionradial}

\subsection{Renormalized Effective Action}

We choose for our field theory model an SU(2) Euclidean
Yang-Mills theory with a complex scalar matter field (in the
fundamental representation), in four dimensional spacetime. As far as our general 
methodology is
concerned, the model choice is not crucial; but, by choosing this
case, we are able to crosscheck readily the findings of the
present work against those of Refs. \cite{idet, wkbpaper}. The case
with Dirac fields is quite similar if one works with the {\it squared} Dirac operator. 
Also, by
considering a gauge theory (rather than the much simpler scalar
field theory), we can demonstrate the {\it gauge invariance} of our
calculational method for the renormalized effective action.

Consider a generic Yang-Mills background : $A_\mu ({\bf x}) \equiv
A_\mu^a ({\bf x}) \frac{\tau^a}{2}$ ($\mu = 1,2,3,4$, $a=1,2,3$;
$\tau$'s denote $2 \times 2$ Pauli matrices). The
Pauli-Villars regularized one-loop effective action associated
with scalar field fluctuations can be represented by
\begin{equation} \label{regularizedeffectiveaction}
\Gamma_\Lambda (A; m) = \ln \left[ \frac{\det (-D^2 + m^2 ) \det
(-\partial^2 + \Lambda^2 )}{\det (-\partial^2 + m^2 ) \det (-D^2 +
\Lambda^2 )} \right]\quad ,
\end{equation}
where $m$ is the scalar mass, $\Lambda$ a heavy regulator mass, and
$D^2$ 
the covariant Laplacian operator
\begin{equation}
D^2 \equiv D_\mu D_\mu \quad , \quad (D_\mu = \partial_\mu - i A_\mu ({\bf
x}) ).
\end{equation}
The Schwinger proper-time representation for the form
(\ref{regularizedeffectiveaction}) is
\begin{equation} \label{effectiveaction-f}
\Gamma_\Lambda (A; m) = - \int_0^\infty \frac{ds}{s} \left(
e^{-m^2 s} - e^{-\Lambda^2 s} \right) F(s)\quad ,
\end{equation}
with $F(s)$ given by  ($s$ is the proper-time variable)
\begin{equation} \label{fdefinition}
F(s) = \int d^4 {\bf x} \; {\rm tr} \left\langle {\bf x} \left|
\left\{ e^{-s (-D^2 )} - e^{-s (-\partial^2)} \right\} \right|
{\bf x} \right\rangle \quad .
\label{fs}
\end{equation}
The proper-time Green's function
\begin{equation}
\Delta({\bf x}, {\bf x^\prime}; s)\equiv \langle {\bf x} | e^{-s (-D^2
)} | {\bf x'} \rangle\quad ,
\label{green}
\end{equation}
admits an asymptotic expansion,  the
DeWitt (or heat kernel) expansion \cite{dewitt, lee}:
\begin{equation} \label{originaldewitt}
\left\langle {\bf x} \left| e^{-s (-D^2 )} \right| {\bf x'}
\right\rangle \sim \frac{1}{(4\pi s)^2} e^{-|{\bf x}-{\bf x'}|^2
/4s} \left\{ \sum_{n=0}^\infty s^n a_n ({\bf x}, {\bf x'})
\right\}\quad, \quad
\textrm{for  } 
s \rightarrow 0+.
\end{equation}
The expansion coefficients, $a_n ({\bf x}, {\bf x'})$ ($n=0,1,2,\cdots$),
and especially the coincidence limits $a_n ({\bf x}, {\bf x})$ of the
first few terms, can be found most simply using  recurrence
relations satisfied by the $a_n ({\bf x}, {\bf x'})$'s. The divergence
structure of $\Gamma_\Lambda (A; m)$ as $\Lambda \rightarrow \infty$
is governed by the values of ${\rm tr} \; a_1 ({\bf x}, {\bf x})$
and ${\rm tr} \; a_2 ({\bf x}, {\bf x})$, and in our case we have \cite{dewitt, lee}
\begin{equation}
{\rm tr} \; a_1 ({\bf x}, {\bf x}) = 0, \qquad {\rm tr} \; a_2 ({\bf
x}, {\bf x}) = -\frac{1}{12} {\rm tr} [F_{\mu\nu} ({\bf x})
F_{\mu\nu} ({\bf x})]\quad ,
\end{equation}
where $F_{\mu\nu} \equiv F_{\mu\nu}^a \frac{\tau^a}{2} = i [D_\mu
, D_\nu ]$ is the field strength. Then the renormalized one-loop effective
action in the minimal subtraction scheme is defined as
\begin{equation} \label{renormalizedeffectiveaction}
\Gamma_{\rm ren}(A; m) = \lim_{\Lambda \rightarrow \infty} \left[
\Gamma_\Lambda (A; m) - \frac{1}{12} \frac{1}{(4\pi)^2} \ln \left(
\frac{\Lambda^2}{\mu^2} \right) \int d^4 {\bf x} \; {\rm tr}
[F_{\mu\nu} ({\bf x}) F_{\mu\nu} ({\bf x}) ] \right]\quad ,
\end{equation}
where $\mu$ is the renormalization scale.

\subsection{Radial Backgrounds}

It is a very difficult problem to explicitly evaluate this renormalized effective action (\ref{renormalizedeffectiveaction}).
For a  generic background gauge field $A_\mu ({\bf x})$, there is currently no known method leading to an {\it exact} evaluation of the one-loop effective action. On the other hand, there are
many interesting physical applications ({\it e.g.}, vortices, monopoles, instantons, \dots) where the gauge background is radially symmetric.  In this paper we show that this radial symmetry is strong enough to permit the computation of the renormalized effective action (\ref{renormalizedeffectiveaction}).

A large class of such radial backgrounds is covered by the ansatz form:
\begin{equation} \label{aform}
A_\mu ({\bf x}) = 2\eta_{\mu\nu a} x_\nu f(r) \frac{\tau^a}{2} +
2\eta_{\mu\nu 3} x_\nu g(r) \frac{\tau^3}{2}\quad , \qquad (r \equiv |{\bf
x}| = \sqrt{x_\mu x_\mu})
\end{equation}
where the radial functions $f(r)$ and $g(r)$ are left unspecified,
and $\eta_{\mu\nu a}$ $(a=1,2,3)$ denote the standard 't Hooft symbols
\cite{thooft}.
With $A_\mu ({\bf x})$ of the form (\ref{aform}), the
covariant Laplacian operator $-D^2$ becomes (here $\frac{\tau^a}{2}
\equiv T_a$)
\begin{eqnarray}
-D^2 &=& -\partial_\mu \partial_\mu + 4i f(r) \eta_{\mu\nu a} T_a
x_\nu \partial_\mu + 4i g(r) \eta_{\mu\nu 3} T_3 x_\nu
\partial_\mu + 4f(r)^2 \eta_{\mu\nu a} \eta_{\mu\lambda b} T_a T_b
x_\nu x_\lambda \nonumber \\
&& + 4g(r)^2 \eta_{\mu\nu 3} \eta_{\mu\lambda 3} T_3^2 x_\nu
x_\lambda + 8 f(r) g(r) \eta_{\mu\nu a} \eta_{\mu\lambda 3} T_a T_3
x_\nu x_\lambda\quad . \label{dsquare}
\end{eqnarray}
We may then define \cite{thooft} the operators $L_a \equiv -\frac{i}{2}
\eta_{\mu\nu a} x_\mu \partial_\nu$ (satisfying angular-momentum
commutation relations $[L_a ,L_b ] = i \epsilon_{abc} L_c$) and
use the relations
\begin{eqnarray}
&& \eta_{\mu\nu a} \eta_{\mu\lambda b} = \delta_{ab}
\delta_{\nu\lambda} + \epsilon_{abc} \eta_{\nu\lambda c}\quad ,
\qquad T_a T_a = \frac{3}{4} {\bf 1}\quad , \nonumber \\
&& -\partial_\mu \partial_\mu = -\frac{\partial^2}{\partial r^2} -
\frac{3}{r} \frac{\partial}{\partial r} + \frac{4}{r^2} {\vec
L}^2\quad , \quad ({\vec L}^2 \equiv L_a L_a )
\end{eqnarray}
to recast the expression (\ref{dsquare}) as
\begin{eqnarray}
-D^2 &=& -\frac{\partial^2}{\partial r^2} - \frac{3}{r}
\frac{\partial}{\partial r} + \frac{4}{r^2} {\vec L}^2 + 8f(r)
\vec T \cdot \vec L + 8g(r) T_3 L_3 \nonumber \\
&& + r^2 \left\{ 3f(r)^2 + g(r)^2 +2f(r)g(r) \right\}\quad .
\label{radialdsquare}
\end{eqnarray}
Based on this form, we may associate an infinite number of
partial-wave radial differential operators with the given system.
We distinguish between three important cases.

\subsubsection{Case 1 : $g(r) \equiv 0$, but $f(r) \neq 0$}

Suppose that $g(r) \equiv 0$, but $f(r) \neq 0$. This is the form relevant to the instanton computation in \cite{idet,thooft}. Then $A_\mu ({\bf x})$ is given by the first piece only on the
right hand side of (\ref{aform}). Then, noting that there
exists another set of angular-momentum-like operators ${\bar L}_a
\equiv -\frac{i}{2} {\bar \eta}_{\mu\nu a} x_\mu \partial_\nu$
(satisfying $[L_a ,{\bar L}_b ] = 0$ and ${\bar L}_a {\bar L}_a =
L_a L_a \equiv {\vec L}^2$ ) \cite{thooft}, partial waves can be
specified by the quantum numbers $J_1 \equiv (l, j, j_3 , {\bar
l}_3 )$, where
\begin{eqnarray}
&& ( {\vec L}^2 )' = l(l+1), \qquad l=0, \frac{1}{2}, 1,
\frac{3}{2}, \cdots \; ; \nonumber \\
&& ( {\vec J}^2 )' = j(j+1), \;\; ({\rm with} \; J_a \equiv L_a
+ T_a ), \; j = \left| l \pm \frac{1}{2} \right| \; ; \nonumber \\
&& ( J_3 )' \equiv j_3 = -j, -j+1, \cdots , j \; ; \nonumber \\
&& ( {\bar L}_3 )' \equiv {\bar l}_3 = -l, -l+1, \cdots , l.
\label{lll}
\end{eqnarray}
The radial differential operator, representing $-D^2$ in the given
partial wave sector, thus assumes the form
\begin{equation} \label{hcase1}
{\cal H}_{J_1} \equiv -D_{(l,j)}^2 = -\partial_{(l)}^2 + 4f(r)
\left[ j(j+1) - l(l+1) - \frac{3}{4} \right] + 3r^2 f(r)^2\quad ,
\end{equation}
where $\partial_{(l)}^2$ is the partial-wave form of the free
Laplacian $\partial_\mu \partial_\mu$ :
\begin{equation} \label{radialfreelaplacian}
\partial_{(l)}^2 \equiv \frac{\partial^2}{\partial r^2} +
\frac{3}{r} \frac{\partial}{\partial r} - \frac{4}{r^2} l(l+1)
\end{equation}

\subsubsection{Case 2: $f(r)=0$, but $g(r) \neq 0$}

The system with $f(r)=0$, but $g(r) \neq 0$, is
simpler: here, partial waves are specified by the quantum numbers
$J_2 \equiv (l, l_3 , t_3 , {\bar l}_3 )$, where $( L_3 )' \equiv
l_3 = -l, -l+1 , \cdots , l$ and $( T_3 )' \equiv t_3 = \pm
\frac{1}{2}$. In this case, the radial differential operator
becomes
\begin{equation} \label{hcase2}
{\cal H}_{J_2} \equiv -D_{(l,l_3 , t_3 )}^2 = -\partial_{(l)}^2 +
8g(r) l_3 t_3 + r^2 g(r)^2 .
\end{equation}

\subsubsection{Case 3: both $f(r)$ and $g(r)$ nonvanishing}

The situation is somewhat more complicated if both $f(r)$ and $g(r)$
are nonvanishing, since $-D^2$ will then be
nondiagonal in either basis considered above. This case can be
treated by allowing partial wave sectors themselves to be
finite-dimensional vector spaces. Explicitly, taking a partial
wave specified by the quantum numbers $J_3 \equiv (l,j_3 , {\bar
l}_3 )$, we can represent the operator $T_3 L_3$ according to
\begin{eqnarray}
T_3 L_3 &\leftrightarrow& \left(
\begin{array}{cc}
\frac{j_3^2}{2l+1} - \frac{1}{4} & -\frac{j_3}{2l+1}
\sqrt{(l+\frac{1}{2})^2 - j_3^2 } \\
-\frac{j_3}{2l+1} \sqrt{(l+\frac{1}{2})^2 - j_3^2 } &
-\frac{j_3^2}{2l+1} - \frac{1}{4}
\end{array}
\right), \; {\rm if} \; j_3 = -l+\frac{1}{2}, \cdots,
l-\frac{1}{2} \nonumber \\
T_3 L_3 &\leftrightarrow& \frac{1}{2} l, \qquad {\rm if} \; j_3 =
\pm (l+\frac{1}{2}),
\end{eqnarray}
where the $2\times 2$ matrix, appearing when $|j_3 | \neq
l+\frac{1}{2}$, is defined relative to the basis vectors $|j=l\pm
\frac{1}{2}\rangle$. This allows us to represent $-D^2$ in the
given partial wave by the (matrix) radial differential operator
\begin{equation} \label{hcase3}
{\cal H}_{J_3} \equiv -D_{(l,j_3 )}^2 =
\begin{cases} { -\partial_{(l)}^2 + W(r) +
Z_{(l,j_3 )} \quad
, \; {\rm if} \; j_3 = -l+\frac{1}{2}, \cdots, l-\frac{1}{2} \cr
-\partial_{(l)}^2 + W(r) +4l f(r) +4l g(r)\quad , \; {\rm if} \; j_3 = \pm
(l+\frac{1}{2}),}
\end{cases}
\end{equation}
where
\begin{eqnarray}
&& W(r) = r^2 \left\{ 3f(r)^2 + g(r)^2 + 2f(r)g(r) \right\}, \\
&& Z_{(l,j_3 )} = \left( {\setlength\arraycolsep{8pt}
\begin{array}{cc} 4l f(r) + 8(\frac{j_3^2}{2l+1}-\frac{1}{4}) g(r) &
-\frac{8j_3}{2l+1} \sqrt{(l+\frac{1}{2})^2 - j_3^2} \; g(r) \\
-\frac{8j_3}{2l+1} \sqrt{(l+\frac{1}{2})^2 - j_3^2} \; g(r) &
-4(l+1)f(r) - 8(\frac{j_3}{2l+1}+\frac{1}{4})g(r)
\end{array}} \right).
\end{eqnarray}
The Gel'fand-Yaglom method has a straightforward generalization 
\cite{kirsten} to matrix-valued operators, so the numerical part 
of the computation follows as before. Such matrix-valued radial 
operators have in fact been considered in \cite{burnier}, and also 
occur naturally when considering fluctuations of a  Dirac-spinor 
matter field in a radially symmetric background.

An interesting subclass of these radial backgrounds consists of those that are {\it self-dual} (or {\it anti-self-dual}). Such gauge fields satisfy automatically the classical
Yang-Mills field equations, and as such they are of particular
importance. With our potential form in (\ref{aform}), such self-dual
or anti-self-dual configurations are obtained if certain special
functional forms are chosen for $f(r)$ and $g(r)$. Explicitly, for
self-dual configurations, the following choices can be made:
\begin{itemize}
\item[(i)] $f(r) = \frac{1}{r^2 + \rho^2}$, $g(r)=0$ (i.e., $A_\mu
= \eta_{\mu\nu a} \frac{x_\nu}{r^2 + \rho^2} \tau^a$) for a single
instanton solution in the regular gauge;
\item[(ii)] $f(r)=0$,
$g(r)=-\frac{B}{2}={\rm const}$. (i.e., $A_\mu = -\eta_{\mu\nu a}
x^\nu \frac{B}{2} \tau^a$) for a uniform self-dual field strength
background;
 \item[(iii)] $f(r)=\frac{b}{ \sinh [b(r^2 + \rho^2
)]}$, $g(r)=b \; \tanh [\frac{b}{2} (r^2 + \rho^2 )]$ for the
so-called Minkowski solution \cite{minkowski} which describes a
single instanton immersed in a uniform background.
\end{itemize}
Note that the Minkowski solution reduces to the case (i) or (ii) in
appropriate limits. Anti-self-dual solutions may be obtained for the
choice $f(r)=\frac{\rho^2}{r^2 (r^2 + \rho^2 )}$, and $g(r)=0$ (i.e.,
$A_\mu = \eta_{\mu\nu a} \frac{x_\nu \rho^2}{r^2 (r^2 + \rho^2 )}$);
this corresponds to a single anti-instanton in the singular gauge.
With any of these classical solutions chosen as the background, our
discussion above tells us that the operator $-D^2$ can be written in
the partial-wave expanded form. In particular, in the case of the
Minkowski solution for which both $f(r)$ and $g(r)$ are
nonvanishing, the related partial-wave differential operator will
take a $2\times 2$ matrix form.

But, in the following analysis, it
will be sufficient to assume that our background potentials are just
of the radial form (\ref{aform}) --- i.e., they {\it do not} have to satisfy
classical field equations.

\subsection{Partial Wave Decomposition of Effective Action}

Taking advantage of this radial symmetry, we can make a partial wave decomposition in (\ref{fs}):
\begin{equation}
F(s)=\sum_J F_J(s)\quad ,
\label{fj}
\end{equation}
where
\begin{equation}
F_J(s)=\int_0^\infty dr\, {\rm tr} \left\{\tilde{\Delta}_J(r, r; s)-\tilde{\Delta}^{\rm free}_J(r, r; s)\right\} \quad .
\label{fjsum}
\end{equation}
The proper-time {\it  radial} Green's function for each partial wave $J$ is defined as
\begin{eqnarray}
\tilde{\Delta}_J(r, r^\prime; s)&\equiv& \langle r | 
e^{-s\, \tilde{{\mathcal H}}_J}\, | r^\prime \rangle
\quad ,
\end{eqnarray}
in terms of the radial operator
\begin{equation}
\tilde{{\mathcal H}}_J
\equiv \frac{1}{r^{3/2}} \mathcal{H}_J\, r^{3/2} 
=-\frac{d^2}{dr^2}+V_J(r)\quad .
\end{equation}
Note that we have extracted a measure factor $r^{3/2}$ in writing
$\tilde{\mathcal H}=r^{-3/2}\,{\mathcal H}\, r^{3/2}$. The form of
the (possibly matrix valued) radial potential $V_J(r)$ depends on
the specific form of the gauge field entering the covariant
Laplacian operator. In each case, the effective radial partial-wave
potential $V_J(r)$
contains 
a centrifugal term,
having the structure 
\begin{equation}
V_J (r) = \frac{4l(l+1)+\frac{3}{4}}{r^2} + U_J (r)\quad ,
\label{centrifugal}
\end{equation}
where $l$ is the half-integer valued quantum number in (\ref{lll}).

The partial-wave-based representation of the regularized effective
action is :
\begin{equation}
\Gamma_\Lambda (A;m) = -\sum_J \int_0^\infty \frac{ds}{s} \left(
e^{-m^2 s} - e^{-\Lambda^2 s} \right) \int_0^\infty dr\, {\rm tr}\,\left\{
\tilde{\Delta}_J (r,r;s) -  \tilde{\Delta}_J^{\rm free} (r,r;s) \right\}\quad .
\end{equation}
Then, for the explicit evaluation of the renormalized effective
action given by (\ref{renormalizedeffectiveaction}), it is
convenient to separate the partial wave sum into two parts
\cite{idet, wkbpaper}: (i)  the sum over partial waves with $J \leq
J_L$ (here, $J_L$ is chosen such that it may refer to some large
$l$-value, $l=L$);  and (ii) the remaining infinite sum involving all
$J>J_L$ terms. In the first contribution involving the finite
$J$-sum, the regulator plays no role in the limit $\Lambda
\rightarrow \infty$, and so may be removed from this sum. Based on this procedure, we can now write
\begin{equation} \label{splitgamma}
\Gamma_{\rm ren} (A;m) = \Gamma_{J \leq J_L} (A;m) + \Gamma_{J
> J_L} (A;m)\quad ,
\end{equation}
with
\begin{eqnarray}
\Gamma_{J \leq J_L} (A;m) &=& -\sum_{J \leq J_L} \int_0^\infty
\frac{ds}{s} e^{-m^2 s} \int_0^\infty dr \; {\rm tr} \left\{
\tilde{\Delta}_J (r,r;s) - \tilde{\Delta}_J^{\rm free} (r,r;s)
\right\} \nonumber \\
&=& \sum_{J \leq J_L} \ln \det \left( \frac{\tilde{\cal H}_J
+m^2}{\tilde{\cal H}_J^{\rm free} +m^2} \right)\quad ,
\label{finitepartgamma} \\
\Gamma_{J>J_L} (A;m) &=& -\sum_{J>J_L} \int_0^\infty \frac{ds}{s}
\left( e^{-m^2 s} - e^{-\Lambda^2 s} \right) F_J (s) \nonumber \\
&& \qquad \qquad - \frac{1}{12} \frac{1}{(4\pi)^2} \ln \left(
\frac{\Lambda^2}{\mu^2} \right) \int d^4 x \; {\rm tr} (F_{\mu\nu}
F_{\mu\nu}). \label{infinitepartgamma}
\end{eqnarray}

In the right hand side of (\ref{finitepartgamma}), $\ln \det (
\frac{\tilde{\mathcal{H}}_J + m^2}{\tilde{\mathcal{H}}_J^{\rm free}
+ m^2} )$ may well be replaced by $\ln \det ( \frac{\mathcal{H}_J +
m^2}{\mathcal{H}_J^{\rm free} + m^2} )$, the two being the same.
In our model cases, we notice that not all quantum numbers in $J$
are relevant for our radial Green function $\Delta_J (r,r';s)$. In
view of this, for the three possible forms of ${\cal H}_J$
considered above [see (\ref{hcase1}), (\ref{hcase2}) and
(\ref{hcase3})], we may express the decomposition formula (\ref{fj})
in the more explicit forms:
\begin{subeqnarray}
& {\it Case \; 1}: \; F(s) &= \sum_{(l,j)} (2j+1)(2l+1) F_{(l,j)}(s)
\nonumber \\
&&= \sum_{l=0,\frac{1}{2},1,\cdots} (2l+1)(2l+2) [ F_{(l,
j=l+\frac{1}{2})} (s) + F_{(l + \frac{1}{2}, j=(l+\frac{1}{2}) -
\frac{1}{2})} (s)]\quad , \quad\slabel{fcase1} \\
& {\it Case \; 2}: \; F(s) &= \sum_{l=0,\frac{1}{2},1,\cdots}
\sum_{l_3 = -l}^l \sum_{t_3 = \pm \frac{1}{2}} (2l+1) F_{(l,l_3 ,
t_3 )}(s) \quad , \slabel{fcase2} \\
& {\it Case \; 3}: \; F(s) &= \sum_{l=0,\frac{1}{2},1,\cdots}
\sum_{j_3 = -(l+\frac{1}{2})}^{l+\frac{1}{2}} (2l+1) F_{(l, j_3
)}(s) \quad . \slabel{fcase3}
\end{subeqnarray}
Note that the degeneracy factors here contain a common factor
$(2l+1)$ from the $\bar{l}_3$-sum.
Here, as the notations of (\ref{fcase1})-(\ref{fcase3}) are used,
the designation $J\leq J_L$ or $J>J_L$ may be identified with the
appropriate division in the values of the quantum number $l$, i.e.,
$J\leq J_L$ when $l\leq L$ ($L$ is some, arbitrarily chosen, large
value) and $J>J_L$ when $l>L$. For example, the low partial wave sum in
(\ref{finitepartgamma}) is expressed explicitly as:
\begin{subeqnarray}
&{\it Case \; 1}:& \nonumber\\
&\Gamma_{J\leq J_L}(A, m) &= \sum_{l=0}^{L} (2l+1)(2l+2)
\{ \ln \frac{ \det (-D_{(l,l+1/2)}^2 +m^2)}{\det(-\partial_{(l)}^2 +m^2)}
+ \ln \frac{\det (-D_{(l+1/2,l)}^2 +m^2)} {\det(-\partial_{(l+1/2)}^2
+m^2 )} \}\nonumber\\
\slabel{pscase1} \\
& {\it Case \; 2}:& \nonumber\\
&\Gamma_{J\leq J_L}(A, m) &= \sum_{l=0}^{L} \sum_{l_3
=-l}^{l} \sum_{t_3 = \pm 1/2} (2l+1) \ln \frac{\det ( -D_{(l,l_3 ,
t_3)}^2 +m^2)} {\det(-\partial_{(l)}^2 +m^2)}
\slabel{pscase2} \\
& {\it Case \; 3}: & \nonumber\\
&\Gamma_{J\leq J_L}(A, m) &= \sum_{l=0}^L
\sum_{j_3 = -(l+\frac{1}{2})}^{l+\frac{1}{2}} (2l+1) \ln \frac{\det ( -D_{(l,j_3)}^2 +m^2)} {\det(-\partial_{(l)}^2 +m^2)}
\slabel{pscases}
\end{subeqnarray}
The low partial wave
contribution,  $\Gamma_{J\leq J_L}$,  in (\ref{finitepartgamma}), may be determined numerically.
On the other hand, the large partial wave contribution, $\Gamma_{J>J_L}$, in (\ref{infinitepartgamma}),
is calculated analytically for large $L$, to the desired accuracy in powers of $1/L$.

\subsection{Low Partial Wave Contribution}

To evaluate the low partial wave contribution $\Gamma_{J\leq J_L}$, we use the Gel'fand-Yaglom technique \cite{gy,levit,coleman,dreyfus,forman,kirsten,kleinert}, which can be summarized as follows.  Suppose ${\cal M}_1$ and ${\cal M}_2$ denote two
second-order radial differential operators
on the interval $r \in [0,\infty)$. Then the ratio of the determinants is given by
\begin{equation} \label{evolve}
\frac{\det\,  {\cal M}_1}{\det\, {\cal M}_2}  = \lim_{R
\rightarrow \infty} \left( \frac{\Phi_1 (R)}{\Phi_2 (R)} \right),
\end{equation}
where $\Phi_i (r)$ $(i=1, 2)$ satisfy the {\it initial value} problems :
\begin{eqnarray}
{\cal M}_i \Phi_i (r) = 0\quad ;
 \quad \Phi_i(r)
\sim r^{2l} \quad{\rm as}\quad  r\to 0\quad .
\end{eqnarray}
Here $l$ is the index in the centrifugal term in (\ref{centrifugal}). Since an initial value problem is extremely simple to solve
numerically, this provides an efficient calculational method for the individual
radial determinants. Here we take ${\cal M}_1 =
{\cal H}_J +m^2$, and ${\cal M}_2 = {\cal H}_J^{\rm free} +m^2$. Thus, $\Phi_2(r)$ is in fact known analytically. Then better  numerical
results are obtained \cite{idet} by
considering directly the initial value problem with the second-order
differential equation derived for the ratio function $S(r) = \ln ( \Phi_1
(r)/ \Phi_2 (r) )$. This method has been implemented successfully in \cite{idet} for the instanton determinant computation, and for the false vacuum decay problem in both flat \cite{fvdecay} and curved \cite{wang} spacetime.
Furthermore, this method of calculating radial
determinants can be generalized to the case when the second-order
differential operator $\cal M$ in question contains a matrix-type
potential \cite{kirsten}, as in our {\it Case 3}. Explicit numerical results for the three radial cases discussed above will be presented in the sequel.

\section{Large Partial-wave Contributions and Renormalization}
\label{sectionlarge}

The large partial wave contribution cannot be evaluated numerically because of the need to remove the heavy mass regulator, $\Lambda$. Instead we compute {\it analytically} the large $L$ behavior of $\Gamma^\Lambda_{J>J_L}$.
To determine the large partial-wave contribution $\Gamma^\Lambda_{J>J_L}$
(given by (\ref{infinitepartgamma})), one needs
the large-$l$ (i.e., $l\gg L$) behavior of the function $F_J (s)$. To
that end, in Refs. \cite{idet, wkbpaper} we used the scattering
phase shift representation of $F_J (s)$, and then the radial WKB
approximation up to second order. Here we present a  much simpler
approach,
introducing a new ``uniform'' DeWitt expansion for the radial
proper-time Green's function $\tilde{\Delta}_J (r,r;s)$, which remains
valid when $l$ becomes large. As we shall see, this new approach
gives rise to results in complete agreement with those from our
earlier method, with much less labor.

\subsection{Uniform DeWitt Expansion}

In the presence of the effective radial potential, $V(r) =\frac{4l(l+1)+\frac{3}{4}}{r^2} + {\cal V}(r)$,
we seek
a large-$l$ asymptotic representation of the related
proper-time Green's function $\tilde{\Delta}(r,r';s)$:
\begin{subeqnarray} \label{green-eqn-boundary}
&& (\partial_s - \partial_r^2 + V(r) ) \tilde{\Delta} (r,r';s) = 0\quad ,
\qquad ({\rm for}
\; s>0) \slabel{green-schequation} \\
&& s \rightarrow 0+: \; \tilde{\Delta} (r,r';s) \longrightarrow
\delta (r-r')\quad . \slabel{green-boundarycondition}
\end{subeqnarray}
We take ${\cal V}(r)$ to be a typical smooth potential. Now,
as $l$ becomes very large, the presence of the large centrifugal
potential, $\frac{4l(l+1)+\frac{3}{4}}{r^2}$, has the
consequence that our Green's function $\tilde{\Delta}(r,r';s)$ acquires a totally negligible
amplitude for $s \gg A/l^2$ (where $A$ is an $O(1)$ constant) . Because of this property, given a certain quantity which
involves the integral of this function over $s$ [such as, for example,
$\Gamma_{J>J_L}(A;m)$, given by (\ref{infinitepartgamma})], it will
suffice to use an accurate representation of
$\tilde{\Delta}(r,r';s)$ for $s$ satisfying the condition $0<s\,l^2
\lesssim O(1)$, and having the property that it becomes
exponentially small for $s \gg A/l^2$. Although only small-$s$
values are relevant, the usual small-$s$ DeWitt expansion [the
one-dimensional analogue of (\ref{originaldewitt})] cannot serve
this purpose since it fails to account for the effect of the large
centrifugal potential term. There is a conflict between the small $s$ limit and the large $l$ limit.
 To see this problem more clearly, consider the behavior of the function $\tilde{\Delta}(r,r';s)$
with ${\cal V}(r)$ set to zero. For this free case, denoted
$\tilde{\Delta}^{\rm free} (r,r';s)$, we have a closed-form expression in
general $n$ spacetime dimensions (see Appendix \ref{appfree}). As $l$
becomes large, this function admits a uniform approximation of the form
(see (\ref{freegreen-uniform2}))
\begin{equation}
\tilde{\Delta}^{\rm free} (r,r';s) \sim \frac{1}{\sqrt{4\pi s}}\,
e^{-\frac{(r-r')^2}{4s} - \frac{4l(l+1)+\frac{3}{4}}{rr'} s} \left\{
1 + O ( s^2 ) \right\}\quad ,
\label{freedewitt}
\end{equation}
valid as long as $s$ is such that $0<s l^2 \lesssim O(1)$. The naive
small-$s$ DeWitt expansion is not adequate for our purpose since it
effectively replaces  the exponential factor
$e^{-\frac{4l(l+1)+\frac{3}{4}}{rr'}s}$ (which can be $O(1)$ for $s
\sim \frac{A}{l^2}$) by the first few terms of its Taylor
series in $s$.

To obtain the desired large-$l$ expansion of our radial proper-time
Green's function, it is convenient to set
\begin{equation}
V(r) = l^2 U(r)\quad ,
\end{equation}
(so that $U(r)$ remains finite as $l\rightarrow
\infty$), and introduce the rescaled proper-time variable
\begin{equation}
t = l^2 s\quad .
\end{equation}
Now the situation for the large-$l$ limit of $\tilde{\Delta}
(r,r';\frac{t}{l^2})$ is actually the same as that appropriate to
the so-called $\frac{1}{\Lambda}$-expansion of the proper-time Green
function considered previously (for a different purpose) in Ref.
\cite{lee2}, identifying $\Lambda$ with $l^2$. Thus, based
on the result of \cite{lee2}, we may immediately write the
$\frac{1}{l}$-expansion of $\tilde{\Delta} (r,r';\frac{t}{l^2})$,
having the structure
\begin{equation} \label{ansatz}
\tilde{\Delta}(r,r'; \frac{t}{l^2} ) = \frac{l}{\sqrt{4\pi t}}
e^{-\frac{l^2 (r-r')^2}{4t}} \left\{ \sum_{k=0}^\infty b_k (r,r';t)
\left( \frac{1}{l^2} \right)^k \right\}\quad ,
\end{equation}
with suitable coefficient functions $b_k (r,r';t)$ which are regular
near $r=r'$. Inserting this form in (\ref{green-schequation}), we
see that the coefficient functions $b_k (r,r';t)$ must satisfy:
\begin{eqnarray}
&& l^2 U(r) \sum_{k=0}^\infty b_k (r,r';t) \left( \frac{1}{l^2}
\right)^k + l^2 \sum_{k=0}^\infty \partial_t b_k (r,r';t) \left(
\frac{1}{l^2} \right)^k \nonumber \\
&& \qquad + \frac{l^2 (r-r')}{t} \sum_{k=0}^\infty \partial_r b_k
(r,r';t) \left( \frac{1}{l^2} \right)^k - \sum_{k=0}^\infty
\partial_r^2 b_k (r,r';t) \left( \frac{1}{l^2} \right)^k = 0.
\label{ansatz-equation}
\end{eqnarray}
We can here regard $U(r)$ to be strictly of order
$(\frac{1}{l^2})^0$, i.e., disregard the fact that it might contain
terms with $\frac{1}{l}$-suppression, to simplify the presentation
of our result. Then, (\ref{ansatz-equation}) gives rise to
recurrence relations satisfied by the coefficient functions $b_k
(r,r';t)$:
\begin{subeqnarray}
O(l^2) &:& U(r) b_0 (r,r';t) + \partial_t b_0 (r,r';t) +
\frac{r-r'}{t} \partial_r b_0 (r,r';t) = 0, \slabel{recursion1} \\
O(l^{2-2k}) &:& U(r) b_k (r,r';t) + \partial_t b_k (r,r';t) +
\frac{r-r'}{t} \partial_r b_k (r,r';t) \nonumber \\
&& \qquad - \partial_r^2 b_{k-1} (r,r';t) = 0, \qquad
(k=1,2,3,\cdots) \slabel{recursion2}
\end{subeqnarray}
Further, because of the boundary condition
(\ref{green-boundarycondition}), we must have $b_k (r,r';t=0) =
\delta_{k 0}$.

To simplify the analysis of the recurrence relations, we may
introduce a new variable $u$ (instead of $r$) by setting $r=r'+t u$
and define a new set of functions,
\begin{equation} \label{newset}
\tilde{b}_k (u,r',t) = e^{\frac{1}{u} \int_{r'}^{r'+t u} U(w) dw}
b_k (r'+tu,r';t).
\end{equation}
[Here we have restricted our attention to the case when $U(r)$ is not a
matrix-valued potential. The case with a matrix-valued potential is
discussed in Appendix \ref{appmatrixdewitt}]. Then the above
recurrence relations can be recast as
\begin{subeqnarray}
O(l^2) &:& \frac{\partial}{\partial t} \tilde{b}_0 (u,r';t) = 0,
\slabel{recursion3} \\
O(l^{2-2k}) &:& \frac{\partial}{\partial t} \tilde{b}_k (u,r';t) =
\frac{1}{t^2} \left\{ \frac{\partial^2}{\partial u^2}
\tilde{b}_{k-1} (u,r';t) - 2 g'(u,r';t) \frac{\partial}{\partial u}
\tilde{b}_{k-1} (u,r';t) \right. \nonumber \\
&& \qquad \left. + \left[ g'(u,r';t)^2 - g''(u,r';t) \right]
\tilde{b}_{k-1} (u,r';t) \right\}, \qquad (k=1,2,\cdots)
\slabel{recursion4}
\end{subeqnarray}
where $g(u,r';t) \equiv \frac{1}{u} \int_{r'}^{r'+tu} U(w) dw$,
$g'(u,r';t) \equiv \frac{\partial}{\partial u} g(u,r';t)$, and
$g''(u,r';t) \equiv \frac{\partial^2}{\partial u^2} g(u,r';t)$.
Since $b_0 (r,r';t=0) = 1$, and so $\tilde{b}_0 (u,r';t=0) = 1$, we
now immediately conclude from (\ref{recursion3}) that $\tilde{b}_0
(u,r';t) = 1$ for any $t>0$. This in turn tells us that
\begin{equation} \label{b0}
b_0 (r,r';t) = e^{-\frac{t}{r-r'} \int_r^{r'} U(w) dw}.
\end{equation}
As a check, we note that if we choose the form $U(r) = \frac{4l(l+1) +
\frac{3}{4}}{l^2 r^2}$ (appropriate to the free case with ${\cal V}(r)=0$), then
(\ref{b0}) reduces to $b_0 (r,r';t) = e^{-\frac{4l(l+1) +
\frac{3}{4}}{rr'} \frac{t}{l^2}}$, producing the correct
exponential factor in (\ref{freedewitt}), related to the centrifugal potential term.
Clearly, in the coincidence limit of $r'=r$, we have
\begin{equation} \label{b0c}
b_0 (r,r;t) = e^{-t U(r)}.
\end{equation}
Using $\tilde{b}_0 (u,r';t) = 1$ in the $k=1$ case of
(\ref{recursion4}), we obtain
\begin{equation} \label{dtb1dt}
\frac{\partial}{\partial t} \tilde{b}_1 (u,r';t) = \frac{1}{t^2}
\left\{ g'(u,r';t)^2 - g''(u,r';t) \right\}.
\end{equation}
Then, to find the coincidence limit of $b_1 (r,r';t)$, i.e., the
expression for $r'=r$, we may set $u=0$ in (\ref{dtb1dt}) (together
with the easily obtained expressions $g'(0,r';t) = \frac{1}{2} t^2
U'(r)$ and $g''(0,r';t) = \frac{1}{3} t^3 U''(r')$) to obtain
\begin{equation}
\frac{\partial}{\partial t} \tilde{b}_1 (u=0,r;t) = \frac{1}{4} t^2
U'(r)^2 - \frac{1}{3} t U''(r).
\end{equation}
This immediately leads to the expression
\begin{equation} \label{b1c}
b_1 (r,r;t) = e^{-t U(r)} \left\{ \frac{1}{12} t^3 U'(r)^2 -
\frac{1}{6} t^2 U''(r) \right\}.
\end{equation}
Higher-order coefficients can be found similarly; for instance, for $b_2
(r,r;t)$ we find
\begin{eqnarray}
b_2 (r,r;t) &=& e^{-t U(r)} \left\{ \frac{1}{288} t^6 U'(r)^4 -
\frac{11}{360} t^5 U'(r)^2U''(r) \right. \nonumber \\
&& \qquad\qquad \left. + \frac{1}{40} t^4 U''(r)^2 + \frac{1}{30}
t^4 U'(r)U^{(3)}(r) - \frac{1}{60} t^3 U^{(4)}(r) \right\}\quad .
\label{b2c}
\end{eqnarray}

We can now  exhibit the desired $\frac{1}{l}$-expansion
structure for our radial proper-time Green's function in the
coincidence limit. Returning to the notations using $V(r)$ and $s$,
it takes, based on the results of (\ref{b0c}) and (\ref{b1c}), the
following form
\begin{equation} \label{green-largel}
\tilde{\Delta}(r,r;s) = \frac{1}{\sqrt{4\pi s}} e^{-s V(r)} \left\{
1 + \left( \frac{1}{12} s^3 V'(r)^2 - \frac{1}{6} s^2 V''(r) \right)
+ O \left( \frac{1}{l^4} \right) \right\}.
\label{mod}
\end{equation}
[See Appendix \ref{appfree} for the explicit verification that this
gives rise to a correct large-$l$ expansion for $\tilde{\Delta}^{\rm free}
(r,r;s)$]. In effect we have resummed all non-derivative terms in the standard DeWitt expansion. These non-derivative terms are all of the form $(s V)^k$ for some $k$, and recalling that $V$ depends quadratically on the partial wave index  $l$, we see that all these terms are of $O(1)$ for $s l^2\sim O(1)$. On the other hand, the remaining terms in the expansion (\ref{mod}) go like $s^3 l^4\sim s (s l^2)^2$, and $s^2 l^2\sim s (s l^2)$, etc, and so remain small in this uniform limit of small $s$ and large $l$.

This modified expansion may be used even when $V(r)$ contains, apart
from $O(l^2 )$ terms, some subleading terms as with the case
$V(r)=l^2 U(r) + l \, T(r) + Q(r)$; in this case, one can also
generate an equally-valid $\frac{1}{l}$-expansion starting from the
form (\ref{green-largel}) by having the exponential of the
subleading terms, i.e., $e^{-\frac{t}{l^2} (l \, T(r) + Q(r))}$
expanded (partly or wholly) in powers of $\frac{1}{l}$. This can be
justified  when $r$ is restricted to the range in which $T(r)$
and $Q(r)$ remain bounded. This trivial
rearrangement can, in fact, be incorporated within our $\frac{1}{l}$-expansion
ansatz (\ref{ansatz}) by allowing the power series development in
the ansatz to have also odd-power terms in $\frac{1}{l}$.

\subsection{Explicit Large Partial Wave Contributions}

We may use the form (\ref{green-largel}), with the formulas (\ref{fjsum}) and
(\ref{infinitepartgamma}), to determine explicitly the large
partial-wave contribution to the effective action. To
facilitate this calculation, we follow Ref. \cite{idet,
wkbpaper} by trading the regulator mass $\Lambda$ for a dimensional
regularization parameter $\epsilon$. This is achieved by demanding
that
\begin{equation} \label{exchange-equation}
-\int_0^\infty \frac{ds}{s} \left( e^{-m^2 s} - e^{-\Lambda^2 s}
\right) F(s) \sim  -\int_0^\infty \frac{ds}{s} e^{-m^2 s} s^{\epsilon}
F(s).
\end{equation}
Since $F(s) = \textrm{(finite constant)} + O(s)$ for small $s$, we
then see that (\ref{exchange-equation}) requires
\begin{equation}
- \ln \left( \frac{\Lambda^2}{m^2} \right) + O \left(
\frac{1}{\Lambda^2} \right) = - \frac{1}{\epsilon} + (\gamma + 2 \ln
m ) + O(\epsilon),
\end{equation}
where $\gamma=0.5772...$ is Euler's constant. Thus the
relation between $\epsilon$ and $\Lambda$ is given by
\begin{equation} \label{regularizationparameters}
\epsilon \longleftrightarrow \frac{1}{\gamma + \ln \Lambda^2}.
\end{equation}
Note that this is only to simplify our calculations; all the
$s$-integrations appearing below can also be carried out within the
original Pauli-Villars regularization framework.

With this preparation, we now proceed to the calculation of
$\Gamma_{J>J_L} (A;m)$ for our {\it Case 1} and {\it Case 2}.
{\it Case 3} will be considered in the sequel.

\subsubsection{Case 1}

With ${\cal H}_{J_1}$ given in (\ref{hcase1}), we have the radial
potential
\begin{equation} \label{vcase1}
V_{(l,j)}(r) = \frac{4l(l+1)+\frac{3}{4}}{r^2} + 4f(r) \left\{
j(j+1) - l(l+1) - \frac{3}{4} \right\} + 3r^2 f(r)^2
\end{equation}
which may be used in (\ref{green-largel}). Then, from
(\ref{fjsum}), $F_J (s)$ for large enough $l$ will follow.
Further, if we represent $F(s)$ by the form (\ref{fcase1}) and use
the correspondence (\ref{regularizationparameters}), it is possible
to express the first part of (\ref{infinitepartgamma}) [the
contribution to $\Gamma_{J>J_L} (A;m)$ other than the
renormalization counterterm] as
\begin{eqnarray}
&& \Gamma_{J>J_L}^\epsilon (A;m) = \int_0^\infty dr \int_0^\infty ds
\left[ -\frac{1}{s} e^{-m^2 s} s^\epsilon
\sum_{l=L+\frac{1}{2}}^\infty
(2l+1)(2l+2) \right. \nonumber \\
&& \quad \left\{ \tilde{\Delta}_{(l,j=l+\frac{1}{2})} (r,r;s) +
\tilde{\Delta}_{(l+\frac{1}{2},j=l)} (r,r;s) - {\tilde
\Delta}_{(l)}^{\rm free} (r,r;s) - {\tilde
\Delta}_{(l+\frac{1}{2})}^{\rm free} (r,r;s) \right\} \Bigg]\quad ,
\label{igcase1}
\end{eqnarray}
where we placed the $s$-integral before the $r$-integral,  in order to give an
explicit result for $\Gamma_{J>J_L} (A;m)$ for general fields. Using
(\ref{green-largel}) in (\ref{igcase1}) with $V_{(l,j)} (r)$ given
by (\ref{vcase1}), the right hand side of (\ref{igcase1}) can be
expressed as
\begin{eqnarray}
&& \Gamma_{J>J_L}^\epsilon (A;m) = \int_0^\infty dr \int_0^\infty ds
\left[ -\frac{1}{s} e^{-m^2 s} s^\epsilon
\sum_{l=L+\frac{1}{2}}^\infty (2l+1)(2l+2) \right. \nonumber \\
&& \quad \frac{1}{\sqrt{4\pi s}} \left\{ e^{-s
V_{(l,l+\frac{1}{2})}(r)} \left( 1 + \left[ \frac{1}{12} s^3
{V_{(l,l+\frac{1}{2})}}'(r)^2 - \frac{1}{6} s^2
{V_{(l,l+\frac{1}{2})}}''(r) \right] + O \left( \frac{1}{l^4}
\right) \right) \right. \nonumber \\
&& \qquad \qquad + e^{-s V_{(l+\frac{1}{2},l)}(r)} \left( 1 + \left[
\frac{1}{12} s^3 {V_{(l+\frac{1}{2},l)}}'(r)^2 - \frac{1}{6} s^2
{V_{(l+\frac{1}{2},l)}}''(r) \right] + O \left( \frac{1}{l^4}
\right) \right) \nonumber \\
&& \qquad \qquad - e^{-s V_{(l)}^{\rm free}(r)} \left( 1 + \left[
\frac{1}{12} s^3 {V_{(l)}^{\rm free}}'(r)^2 - \frac{1}{6} s^2
{V_{(l)}^{\rm free}}''(r) \right] + O \left( \frac{1}{l^4}
\right) \right) \nonumber \\
&& \left. \qquad \qquad - e^{-s V_{(l+\frac{1}{2})}^{\rm free}(r)} \left( 1 +
\left[ \frac{1}{12} s^3 {V_{(l+\frac{1}{2})}^{\rm free}}'(r)^2 - \frac{1}{6}
s^2 {V_{(l+\frac{1}{2})}^{\rm free}}''(r) \right] + O \left( \frac{1}{l^4}
\right) \right) \right\} \Bigg]\quad ,\quad \label{igcase1explicit}
\end{eqnarray}
where $V_{(l)}^{\rm free}(r) \equiv \frac{4l(l+1)+\frac{3}{4}}{r^2}$. We
remark that if we consider the total of all explicitly-kept terms in
the integrand of (\ref{igcase1explicit}), the neglected terms would
at most be $O(\frac{1}{l^5})$; this happens because the leading,
i.e., order-$\frac{1}{l^4}$ terms coming from the four pieces
denoted $O(\frac{1}{l^4})$ in (\ref{igcase1explicit}) (which are
given in terms of $b_2(r,r;t)$) necessarily cancel, as the leading
terms of the potential $V$ match those of $V^{\rm free}$. The above
expression is fully equivalent to that found using the 2nd-order
radial WKB approximation for phase shifts in Refs. \cite{idet,
wkbpaper}. [Actually, in Refs. \cite{idet, wkbpaper}, the WKB
approximation was used with the Langer-modified radial potential
\cite{langer}; but this is inessential for large partial-wave
contributions as the difference corresponds to a trivial
rearrangement of our $\frac{1}{l}$-expansion series]. It is not
difficult to see that this equivalence between the result based on
the radial WKB approximation and our present approach using the
$\frac{1}{l}$-expansion persists to even higher orders also. But
our new $\frac{1}{l}$-expansion for the radial proper-time
Green function is considerably simpler.

The $l$-sum in (\ref{igcase1explicit}) can be performed with the
help of the Euler-Maclaurin summation formula
\begin{equation} \label{euler-maclaurin}
\sum_{l=L+\frac{1}{2}}^\infty f(l) = 2 \int_L^\infty dl \; f(l) -
\frac{1}{2} f(L) - \frac{1}{24} f'(L) + \cdots.
\end{equation}
All terms in this expansion, including the integral term, can be
computed analytically. The result of this calculation, which is
rather lengthy, is given in Appendix \ref{appeuler}. We thus obtain
an explicit double-integral representation for
$\Gamma_{J>J_L}^\epsilon (A;m)$ (see the expressions given in
(\ref{suma})-(\ref{appeulerresult})). Then the integration over the
proper-time variable $s$ can be performed in a straightforward
manner. After carrying out these $s$-integrations, we find that the
quantity $\Gamma_{J>J_L}^\epsilon (A;m)$ for sufficiently large $L$
is given explicitly by the form
\begin{eqnarray}
&& \Gamma_{J>J_L}^\epsilon (A;m) = \frac{1}{8\epsilon} \int_0^\infty
dr \; r^3 \left[ 4h(r)^2 + ( 2f(r) + r f'(r) )^2 \right] \nonumber
\\
&& + \int_0^\infty dr \Bigg[ - 2 r (L+2) h(r) \sqrt{4L^2 +m^2 r^2 }
- \frac{r^3}{8} \left\{ 2 \ln \left( \frac{4L}{r} \right) + \gamma
\right\} \left\{ 4h(r)^2 + (2f(r)+r f'(r))^2 \right\}
\nonumber \\
&& + \frac{L r}{12\sqrt{4L^2 + m^2 r^2}} \left\{ h(r) (24 m^2 r^2 +
44 h(r) r^2 -75 ) + 39 f(r)^2 r^2 + (6 r^2 f'(r)^2 +f''(r) )
r^2 \right. \nonumber \\
&& \quad \left. + f(r)( -2f''(r)r^4 + 24 f'(r) r^3 -9 ) \right\}
\Bigg] + O \left( \frac{1}{L} \right) + O(\epsilon),
\label{case1regularized}
\end{eqnarray}
where $h(r) \equiv f(r) [r^2 f(r) -1]$.

For the expression of $\Gamma_{J>J_L} (A;m)$ we must subtract the
renormalization counterterm from the expression
(\ref{case1regularized}). Using (\ref{regularizationparameters}),
the renormalization counterterm appearing in the definition of the renormalized effective action (\ref{renormalizedeffectiveaction}) is
\begin{equation} \label{counterterm}
\frac{1}{12} \frac{1}{(4\pi)^2} \left( \frac{1}{\epsilon} - \gamma -
\ln \mu^2 \right) \int d^4 x \; {\rm tr} (F_{\mu\nu}F_{\mu\nu})\quad .
\end{equation}
In this radial ansatz  case, $F_{\mu\nu}$ is equal to
\begin{equation}
F_{\mu\nu} = -2 \eta_{\mu\nu a} f(r)(r^2 f(r)- 1) \tau^a +
\frac{x_\lambda}{r} (x_\mu \eta_{\nu\lambda a} - x_\nu
\eta_{\mu\lambda a} )(2r f(r)^2 + f'(r)) \tau^a.
\end{equation}
Hence the counterterm reads
\begin{equation} \label{case1counterterm}
\frac{1}{8} \left( \frac{1}{\epsilon} - \gamma - \ln \mu^2 \right)
\int_0^\infty dr \; r^3 \left[ 4r^4 f(r)^4 - 8r^2 f(r)^3 + 8f(r)^2
+4r f'(r) f(r) +r^2 f'(r)^2 \right].
\end{equation}
Comparing the expression (\ref{case1counterterm}) with
(\ref{case1regularized}), we see that the divergence terms as $\epsilon\to 0$ match precisely  between the two. This also verifies that our
renormalization procedure is a {\it gauge-invariant} one. The full
large partial-wave contribution to the effective action, the sum of
the expression in (\ref{case1regularized}) and minus the result in
(\ref{case1counterterm}), now becomes
\begin{eqnarray}
&& \Gamma_{J>J_L} (A;m) = \int_0^\infty dr \left[ \frac{r^3}{4} \ln
\left( \frac{\mu r}{4L} \right) \left\{ 4h(r)^2 + ( 2f(r)
+ r f'(r) )^2 \right\} \right. \nonumber \\
&& \qquad - 2 r (L+2) h(r) \sqrt{4L^2 +m^2 r^2 } + \frac{L
r}{12\sqrt{4L^2 + m^2 r^2}} \left\{ h(r) ( 24 m^2 r^2 + 44 h(r) r^2
-75 ) \right. \nonumber \\
&& \qquad \quad  \left. + 39 f(r)^2 r^2 + (6 r^2 f'(r)^2 +f''(r) )
r^2 + f(r)( -2f''(r)r^4 + 24 f'(r) r^3 -9 ) \right\} \Bigg]
\nonumber \\
&& \qquad + O \left( \frac{1}{L} \right). \label{case1final}
\end{eqnarray}
This generalizes the result of Refs. \cite{idet, wkbpaper} where the
calculation was performed assuming the special form $f(r) =
\frac{1}{r^2+\rho^2}$ (i.e., the single instanton solution).

For any given radial function $f(r)$, one can then consider the sum
of this analytic expression (\ref{case1final}) for the large partial-wave contribution,
and the numerically determined result for $\Gamma_{J \leq J_L}
(A;m)$ (based on (\ref{finitepartgamma}) and the relation
(\ref{evolve})) to determine the corresponding full renormalized
effective action. Each has quadratic, linear and log divergences for large $L$.
For large $L$, the
$L$-dependence in the two expressions  cancels, as was originally found in \cite{idet}. In fact,
to improve the numerical efficiency of the full effective action
calculation, one may  extend the large $L$ expression for
$\Gamma_{J>J_L} (A;m)$ in (\ref{case1final}), to include also terms up to $O(\frac{1}{L^2})$.
This can  be done with the help of our $\frac{1}{l}$-expansion, using the explicit expression for $b_2
(r,r;t)$ in (\ref{b2c}).

\subsubsection{Case 2}

With ${\cal H}_{J_2}$ given in (\ref{hcase2}), we have the radial
potential
\begin{equation} \label{vcase2}
V_{(l,l_3 , t_3 )}(r) = \frac{4l(l+1)+\frac{3}{4}}{r^2} + 8g(r) l_3
t_3 + r^2 g(r)^2.
\end{equation}
If we represent $F(s)$ by the form (\ref{fcase2}) and use the
correspondence (\ref{regularizationparameters}), the first part of
(\ref{infinitepartgamma}) can be expressed by
\begin{eqnarray}
\Gamma_{J>J_L}^\epsilon (A;m) &=& \int_0^\infty dr \int_0^\infty ds
\Bigg[ -\frac{1}{s} e^{-m^2 s} s^\epsilon \nonumber \\
&& \left. \sum_{l=L+\frac{1}{2}}^\infty (2l+1) \sum_{l_3 = -l}^l
\sum_{t_3 = \pm \frac{1}{2}} \left\{ \tilde{\Delta}_{(l, l_3 , t_3
)} (r,r;s) - \tilde{\Delta}_{(l)}^{\rm free} (r,r;s) \right\} \right].
\label{igcase2}
\end{eqnarray}
For the function $\tilde{\Delta}_{(l, l_3 , t_3 )} (r,r;s)$
(or $\tilde{\Delta}_{(l)}^{\rm free} (r,r;s)$) on the right hand side, we
may use the $\frac{1}{l}$-expansion result in (\ref{green-largel})
with $V(r)$ taken to be equal to the radial potential in
(\ref{vcase2}) (the radial potential $V_{(l)}^{\rm free} (r) =
\frac{4l(l+1)+\frac{3}{4}}{r^2}$). The $l_3$-sum and $t_3$-sum can
be done explicitly, using the formulas :
\begin{subeqnarray}
&& \sum_{l_3 =-l}^l \sum_{t_3 = \pm \frac{1}{2}} e^{-8 l_3 t_3 g(r)
s} = \frac{2 \{ e^{4(l+1)g(r)s} - e^{-4lg(r)s} \}}{e^{4g(r)s}-1},
\slabel{l3sum1} \\
&& \sum_{l_3 =-l}^l \sum_{t_3 = \pm \frac{1}{2}} l_3 t_3 e^{-8 l_3
t_3 g(r) s} = \frac{1}{(e^{4g(r)s}-1)^2} \left\{ l e^{-4lg(r)s} - l
e^{4(l+2)g(r)s} \right. \nonumber \\
&& \qquad\qquad\qquad\qquad\qquad\qquad \left. +
(l+1)e^{4(l+1)g(r)s} -(l+1)e^{-4(l-1)g(r)s} \right\}.
\end{subeqnarray}
We then perform the $l$-sum with the help of the Euler-Maclaurin
summation formula; this produces a double-integral representation in
which the integration over the proper-time variable $s$ can be
executed without too much difficulty. The result of these
manipulations is
\begin{eqnarray}
&& \Gamma_{J>J_L}^\epsilon (A;m) = \frac{1}{24\epsilon}
\int_0^\infty dr \; r^3 \left[ 8g(r)^2 + 4r g'(r)g(r) + r^2 g'(r)^2
\right] \nonumber \\
&& + \int_0^\infty dr \left[ - r^3 g(r)^2 \sqrt{4L^2 +m^2 r^2 } -
\frac{r^3}{24} \left\{ 2 \ln \left( \frac{4L}{r} \right) + \gamma
\right\} \left\{ 8g(r)^2 + 4r
g'(r)g(r) + r^2 g(r)^2 \right\} \right. \nonumber \\
&& + \frac{L r^3}{90\sqrt{4L^2 + m^2 r^2}} \left\{ 6r^4 g(r)^4 +
(20-240L) g(r)^2 -5r ( r g''(r) -12g'(r) ) g(r) + 15r^2
g'(r)^2 \right\} \Bigg] \nonumber \\
&& + O \left( \frac{1}{L} \right) + O ( \epsilon ).
\label{case2regularized}
\end{eqnarray}
The field strength appropriate to this {\it Case 2} reads
\begin{equation}
F_{\mu\nu} = -2 \eta_{\mu\nu a} g(r) \tau^3 + \frac{x_\lambda}{r}
(x_\mu \eta_{\nu\lambda 3} - x_\nu \eta_{\mu\lambda 3} ) g'(r)
\tau^3,
\end{equation}
and hence the renormalization counterterm (\ref{counterterm})
is given by
\begin{equation} \label{case2counterterm}
\frac{1}{24} \left( \frac{1}{\epsilon} - \gamma - \ln \mu^2 \right)
\int_0^\infty dr \; r^3 \left[ 8g(r)^2 + 4r g'(r)g(r) + r^2 g'(r)^2
\right].
\end{equation}
Again we see that the divergent terms as $\epsilon \rightarrow 0$ in
(\ref{case2regularized}) match precisely those of the
renormalization counterterm. The full large-partial wave
contribution, including the renormalization term, is thus given by
\begin{eqnarray}
&& \Gamma_{J>J_L} (A;m) = \int_0^\infty dr \left[ \frac{r^3}{12} \ln
\left( \frac{\mu r}{4L} \right) \left\{ 8g(r)^2 + 4r g'(r)g(r) + r^2
g'(r)^2 \right\} \right. \nonumber \\
&& \qquad - r^3 g(r)^2 \sqrt{4L^2 +m^2 r^2 } + \frac{L
r^3}{90\sqrt{4L^2 + m^2 r^2}} \left\{ 6r^4 g(r)^4 + (20-240L^2 )
g(r)^2 \right. \nonumber \\
&& \qquad \quad \left. -5r ( r g''(r) -12g'(r) ) g(r) + 15r^2
g'(r)^2 \right\} \Bigg] + O \left( \frac{1}{L} \right).
\label{case2final}
\end{eqnarray}
This expression can now be combined with the numerical low partial wave contribution to determine the finite renormalized
effective action in a  radial Yang-Mills background of the form
$A_\mu (x) = 2\eta_{\mu\nu 3} x_\nu g(r) \frac{\tau^3}{2}$.

\section{Cases with Asymptotically Uniform Field Strengths}
\label{constantsection}

In Ref. \cite{lee2} the basis for the validity of the
$\frac{1}{l}$-expansion structure (\ref{ansatz}) was a perturbative
argument: i.e., in the effective potential $V(r) = \frac{4l(l+1)
+ \frac{3}{4}}{r^2} + {\cal V}(r)$, ${\cal V}(r)$ can be treated as
a perturbation to the centrifugal potential term. We then saw in the
previous section that the resulting explicit expression in
(\ref{green-largel}), if used to calculate the large partial-wave
contribution of the effective action, yields a result completely
equivalent to that obtained from using the quantum mechanical WKB
approximation with scattering phase shifts \cite{idet,wkbpaper}.
It is not immediately clear what happens if the potential is unbounded as 
$r\to\infty$, in which case in the WKB language one should use a bound state 
analysis rather than a scattering analysis. This is precisely the potential form 
that arises when the gauge field strength approaches a nonzero constant value at infinity.
Indeed, in our {\it Case 2} with $g(r) =
-\frac{B}{2}$ (leading to uniform self-dual field strengths), we
have the quadratic potential (see (\ref{vcase2}))
\begin{equation} \label{constantcalv}
{\cal V}_{(l,l_3,t_3)} (r) = -4Bl_3t_3 + \frac{B^2}{4}r^2 \quad .
\end{equation}
Thus, the eigenstates of the corresponding radial operator ${\cal
H}_{J_2}$ consist of only bound states. In principle, for the large
partial-wave contribution to the effective action, one might still
try to use the quantum mechanical WKB approximation; but, the WKB
approximation for bound states has a rather different structure from
that for scattering states.

It is thus an important  issue to know whether or not the
$\frac{1}{l}$-expansion for the radial proper-time Green function,
considered in the previous section, retains its validity even
when the potential ${\cal V}(r)$ blows up for large $r$, as above.
In fact,  our $\frac{1}{l}$-expansion-based formula in
(\ref{green-largel}) {\it does} describe the correct asymptotic form valid for
all $r \in (0,\infty)$ even with such an unbounded potential; with the proviso that in (\ref{green-largel}),  the
exponential prefactor $e^{-sV(r)}$ cannot be replaced by its (truncated) power
series in $s$, for the form to be valid even for very large $r$.
That is, we also have {\it infrared} physics captured correctly by
our large-$l$ limit form of the radial proper-time Green function.
We demonstrate this below through the treatment of an important special
case, that of uniform self-dual field strengths for which the
quadratic potential (\ref{constantcalv}) is relevant. Note that the
effective action in a uniform self-dual field strength background
has been studied by many authors before \cite{leutwyler,flory,dunneschubert}, but not using
the partial-wave-based proper-time formalism. This
analysis will also provide a useful comparison when we consider, in
the sequel to this paper, the effective action in the inhomogeneous
background described by the form (\ref{aform}) with $f(r)=0$ and
$g(r)=-\frac{B}{2} \tanh (r/r_0)$.

In our procedure the calculation of the one-loop effective action in
the gauge field background $A_\mu ({\bf x}) = -\eta_{\mu\nu a}x^\nu
\frac{B}{2}\tau^a$, starts from the study of the radial
proper-time Green function $\tilde{\Delta}_{(l,l_3,t_3)} (r,r';s)$,
satisfying the equation
\begin{equation} \label{constantgreenequation}
\left( \partial_s - \partial_r^2 + \frac{4l(l+1)+\frac{3}{4}}{r^2} -
4Bl_3t_3 + \frac{B^2}{4}r^2 \right) \tilde{\Delta}_{(l,l_3,t_3)}
(r,r';s) = 0, \quad (s>0).
\end{equation}
Just like the free radial proper-time Green function discussed in
Appendix {\ref{appfree}, in this case it is possible to find the
corresponding radial Green function in a closed form,  by using (for example)
the method of quantum canonical transformations \cite{leyvraz,
tsaur}. It is given by an expression involving the modified Bessel
function
\begin{equation} \label{constantgreen}
\tilde{\Delta}_{(l,l_3,t_3)} (r,r';s) = \frac{B \sqrt{rr'}}{2\sinh
(Bs)} e^{-\frac{1}{4} B \coth(Bs) (r^2+r'^2) + 4Bl_3t_3s} I_{2l+1}
\left( \frac{Brr'}{2\sinh (Bs)} \right).
\end{equation}
We check the
large-$l$ limit of the expression (\ref{constantgreen}) directly
against our general formula (\ref{green-largel}), as the potential is
specialized to the form $V(r) = \frac{4l(l+1) + \frac{3}{4}}{r^2} -
4Bl_3t_3 + \frac{B^2}{4}r^2$. As we explained earlier, we want our
large-$l$ limit expression to have validity for $s$ satisfying the
condition $0<sl^2 \lesssim O(1)$. But no restriction follows on the
range of the radial coordinate $r$, and so, for the given potential
(which diverges quadratically for large $r$), we want our large-$l$
limit form to be faithful with the true behavior for arbitrarily
large value of $r^2s$. With this point kept in mind, we may set
$s=t/l^2$ and use the uniform asymptotic expansion for large orders
for the modified Bessel function in Appendix \ref{appfree} (see
(\ref{bessel-uniformasymptotic})) with our expression
(\ref{constantgreen}), in the coincident limit,  to obtain the large-$l$ limit form
\begin{eqnarray}
\tilde{\Delta}_{(l,l_3,t_3)} \left( r,r;\frac{t}{l^2} \right) &=&
\frac{Br}{2\sinh \left( \frac{Bt}{l^2} \right)}
\frac{e^{-\frac{1}{2} Br^2 \coth \left( \frac{Bt}{l^2} \right) +
4Bl_3t_3 \frac{t}{l^2} + \nu \left( \sqrt{1+\tilde{z}^2} + \ln
\frac{\tilde{z}}{\sqrt{1+\tilde{z}^2}+1} \right)}}
{\sqrt{2\pi\nu}(1+\tilde{z}^2)^{1/4}} \nonumber \\
&& \qquad \times \left( 1 + \frac{3x-5x^3}{24\nu} + O \left(
\frac{1}{\nu^2} \right) \right), \label{complexasymptotic}
\end{eqnarray}
where $\nu=2l+1$, $\tilde{z} \equiv \frac{Br^2}{2\nu\sinh (
\frac{Bt}{l^2} )}$, and $x \equiv \frac{1}{\sqrt{1+\tilde{z}^2}}$.

For $l$ very large and $0<t\lesssim O(1)$ (but with no restriction
on the value of $r$), it is possible to simplify the complicated
exponential term in (\ref{complexasymptotic}) by keeping in its
exponent only the piece $-\frac{4t}{r^2} - \frac{1}{4}B^2r^2
\frac{t}{l^2}$, namely the leading terms of the given exponent when $l$
becomes very large (but for an unrestricted value of $r$), and
expand all the other terms in increasing powers of $\frac{1}{l}$.
Since $\tilde{z}$ is $O(l)$, this also implies expansion in powers
of $\frac{1}{\tilde{z}}$. Then, from (\ref{complexasymptotic}), we
obtain the following expansion
\begin{eqnarray}
&& \tilde{\Delta}_{(l,l_3,t_3)} \left( r,r;\frac{t}{l^2} \right)
\nonumber \\
&& \qquad = \frac{l e^{-\frac{4t}{r^2} - B^2r^2
\frac{t}{4l^2}}}{\sqrt{4\pi t}} \left\{ 1 - \frac{4t}{r^2}
\frac{1}{l} + \left( \frac{16t^3}{3r^6} + \frac{4t^2}{r^4} +
4Bl_3t_3t - \frac{3t}{4r^2} \right) \frac{1}{l^2} + O \left(
\frac{1}{l^3} \right) \right\}. \label{constantexplicitseries}
\end{eqnarray}
It should be noted that, in the same limit, the leading terms of our
potential, 
$V(r) = \frac{4l(l+1)+\frac{3}{4}}{r^2} - 4Bl_3t_3 +
\frac{B^2}{4}r^2$, are given by $\frac{4l^2}{r^2} +
\frac{B^2}{4}r^2$. Now, according to the same kind of reasoning as
discussed after (\ref{green-largel}), we may replace
(\ref{constantexplicitseries}) by another expansion in which the
exponential factor at front is assumed by $e^{-V(r)\frac{t}{l^2}}$;
this rearrangement is harmless for arbitrarily large values of
$r^2\frac{t}{l^2}$ here. The result of this rearrangement is to make
(\ref{constantexplicitseries}) turn into the structure predicted by
our $\frac{1}{l}$-expansion formula (\ref{green-largel}), with all
factors precisely equal.

We will now show that our radial proper-time Green function
(\ref{constantgreen}) can be used to rederive the known
expression for the effective action. In a uniform self-dual field strength background, it is
known that \cite{flory,dunneschubert}
\begin{equation} \label{constantcgreen}
{\rm tr} \langle {\bf x} s| {\bf x} \rangle = \frac{2}{(4\pi s)^2}
\frac{(Bs)^2}{\sinh^2 (Bs)}.
\end{equation}
Thus we have, as this form is used with
(\ref{effectiveaction-f}), (\ref{fdefinition}) and
(\ref{renormalizedeffectiveaction}),
\begin{eqnarray}
\Gamma_{\rm ren} (A;m) &=& - \frac{1}{(4\pi)^2} \frac{2}{3}B^2 \ln
\left( \frac{m^2}{\mu^2} \right) \nonumber \\
&& -2\int d^4 {\bf x} \int_0^\infty \frac{ds}{s} e^{-m^2s}
\frac{1}{(4\pi s)^2} \left[ \frac{(Bs)^2}{\sinh^2 (Bs)} - 1 +
\frac{1}{3}(Bs)^2 \right].
\end{eqnarray}
Here, in view of (\ref{fdefinition}) and
(\ref{fjsum}), it will suffice to show that
\begin{equation} \label{anglesumgoal}
\sum_{l=0,\frac{1}{2},1,\cdots} \sum_{l_3=-l}^l
\sum_{t_3=\pm\frac{1}{2}} (2l+1) \tilde{\Delta}_{(l,l_3,t_3)}
(r,r;s) = 2\pi^2r^3 \; {\rm tr} \langle {\bf x} s| {\bf x} \rangle
\end{equation}
since $d^4 {\bf x} = 2\pi^2r^3 dr$ after the angular integration.
Inserting the form (\ref{constantgreen}) for
$\tilde{\Delta}_{(l,l_3,t_3)} (r,r;s)$ into the left hand side of
(\ref{anglesumgoal}) and carrying out the $l_3$ and $t_3$ sums
yields the expression
\begin{eqnarray}
&& \sum_{l=0,\frac{1}{2},1,\cdots} \frac{(2l+1)Br}{\sinh^2 (Bs)}
\sinh ((2l+1)Bs) e^{-\frac{1}{2} Br^2 \coth (Bs)} I_{2l+1} \left(
\frac{Br^2}{2\sinh (Bs)} \right)
\nonumber \\
&& \qquad = \frac{Br}{\sinh^2 (Bs)} e^{-\frac{1}{2} Br^2 \coth (Bs)}
\sum_{\nu=1}^\infty \sum_{k=0}^\infty \nu \sinh (\nu Bs)
\frac{z^{2k+\nu}}{\Gamma(k+\nu+1)k!}\quad , \label{constantanglesum}
\end{eqnarray}
where we set $\nu=2l+1$ and $z=\frac{Br^2}{4\sinh (Bs)}$, aside from
using the power series representation of the modified Bessel
function. If we change the summation over $(k,\nu)$ to those over
$(k,n=2k+\nu)$ and use the relation
\begin{eqnarray}
&& \sum_{\nu=1}^\infty \sum_{k=0}^\infty \nu \sinh (\nu Bs)
\frac{z^{2k+\nu}}{\Gamma(k+\nu+1)k!} = \sum_{n=1}^\infty z^n
\sum_{k=0}^{k_m} \frac{(n-2k) \sinh ((n-2k)Bs)}{\Gamma(n-k+1)k!}
\nonumber \\
&& \qquad = \sum_{n=1}^\infty z^n \frac{2^{n-1} \sinh(Bs)
\cosh^{n-1}(Bs)}{(n-1)!} \nonumber \\
&& \qquad = \frac{1}{4}Br^2 \sum_{n=1}^\infty \frac{ \left[
\frac{1}{2}Br^2 \coth(Bs) \right]^{n-1}}{(n-1)!} \nonumber \\
&& \qquad = \frac{1}{4}Br^2 e^{\frac{1}{2}Br^2 \coth(Bs)}
\end{eqnarray}
(here $k_m = \frac{n-1}{2} \; (\frac{n}{2})$ if $n$ is odd (even)),
we then see that the expression in (\ref{constantanglesum}) reduces
to
\begin{equation}
\frac{B^2r^3}{4\sinh^2(Bs)}.
\end{equation}
This coincides with the expression for $2\pi^2r^3 \; {\rm tr}
\langle {\bf x} s| {\bf x} \rangle$ when (\ref{constantcgreen}) is
used for ${\rm tr} \langle {\bf x} s| {\bf x} \rangle$. Hence we
have the relation (\ref{anglesumgoal}) established.

\section{Conclusions} \label{discussion}

In this work we have simplified significantly the calculational
method for the one-loop effective action developed in Refs.
\cite{idet, wkbpaper}, so that any radially symmetric background
case may now be studied with calculational efficiency. The computation is split into two parts: the contribution from low partial waves is calculated numerically using the Gel'fand-Yaglom technique, and the contribution from high partial waves has been computed analytically using a modified DeWitt expansion. It is no longer
necessary to invoke the results of higher-order quantum-mechanical
WKB approximation explicitly --- this is now automatically accounted
for by using the $\frac{1}{l}$-expansion for the radial proper-time
Green function. The main results are contained in the expressions (\ref{case1final}) and (\ref{case2final}) for the analytic behavior of the large partial wave contribution to the renormalized effective action, for two general classes of radially symmetric gauge fields.
Our approach observes gauge invariance, and can be used for any mass value for the associated quantum
fluctuations. It can also be applied to the case with nonvanishing
asymptotic backgrounds. In the sequel, we shall report an extensive
analysis of the Yang-Mills one-loop effective action (not only for
scalar matter but also for fermion fields as well), taking the
radial gauge-field background form of the present work. We can then
use these results to check for instance the range of validity of the
derivative expansion \cite{dunneschubert,wkbpaper}.

In this work we have used the $\frac{1}{l}$-expansion to
calculate the large partial-wave contribution to the effective
action. This expansion could alternatively be used to calculate
{\it approximately} the lower partial-wave contributions as well. Aside
from the Langer modification \cite{langer} which can easily be
incorporated in our $\frac{1}{l}$-expansion, this is effectively
what we have done in Ref. \cite{wkbpaper} with the Yang-Mills
instanton background; there, the instanton determinant was
found to be good to 5\% accuracy. One might be somewhat surprised by
this success. But it need not be so surprising; observe that the
$\frac{1}{l}$-expansion as given in (\ref{green-largel}) also serves
to generate a systematic derivative expansion. In fact, using the
expansion (\ref{green-largel}), we have studied several cases of
one-dimensional functional determinants (including power-like
potentials and the case of $V(x) \propto {\rm sech}^2 x$), to find
that the deviation from the exact value is typically not more than
5\%. This observation could potentially be used to obtain simple approximate estimates for general radial background fields.

\vskip .5cm {\bf Acknowledgments:} 
GD acknowledges the support of the US DOE through grant
DE-FG-02-92ER40716, and thanks the CSSM and Adelaide University for
hospitality and support.
The work of JH and CL was supported by the Korea Research Foundation
Grant funded by Korean Government (MOEHRD) (R14-2003-012-01002-0).

\newpage
\appendix

\section{The Free Radial Proper-time Green Function in General
Spacetime Dimension} \label{appfree}

In this appendix we find the explicit form of the free radial proper-time
Green's function in $n$ spacetime dimension and then discuss its large
angular-momentum limit (to facilitate the application of our
approach in problems with spacetime dimension not equal to four). In
$n$ dimensions, the consideration of the Laplacian $\partial_\mu
\partial_\mu$ in generalized spherical coordinates leads to the
radial differential operator
\begin{equation}
\partial_{\kappa/2}^2 = \frac{\partial^2}{\partial r^2} +
\frac{n-1}{r} \frac{\partial}{\partial r} -
\frac{\kappa(\kappa+n-2)}{r^2},
\end{equation}
where $r=\sqrt{x_1^2 + \cdots + x_n^2}$, and $\kappa = 0,1,2,\cdots$.
[From this form, our expression (\ref{radialfreelaplacian}) is
recovered upon setting $n=4$ and $\kappa=2l$]. Then, noting that
$d^n x = r^{n-1} d^{n-1} \Omega$, we require the free radial proper-time
Green's function
$\Delta_{\kappa/2}^{\rm free} (r,r';s)$ 
to satisfy the
conditions
\begin{subeqnarray}
&& \left\{ \frac{\partial}{\partial s} - \frac{\partial^2}{\partial
r^2} - \frac{n-1}{r} \frac{\partial}{\partial r} +
\frac{\kappa(\kappa+n-2)}{r^2} \right\} \Delta_{\kappa/2}^{\rm free}
(r,r';s)
= 0, \; ({\rm for} \; s>0) \slabel{general-green-schequation} \\
&& s \rightarrow 0+: \; \Delta_{\kappa/2}^{\rm free} (r,r';s)
\longrightarrow \frac{1}{r^{n-1}} \delta (r-r').
\slabel{general-green-boundarycondition}
\end{subeqnarray}
We introduce the modified radial proper-time Green's function
$\tilde{\Delta}_{\kappa/2}^{\rm free} (r,r';s)$ according to
\begin{equation}
\tilde{\Delta}_{\kappa/2}^{\rm free} (r,r';s) = r^{\frac{n-1}{2}}
\Delta_{\kappa/2}^{\rm free} (r,r';s) r'^{\frac{n-1}{2}}.
\end{equation}
In terms of this function, (\ref{general-green-schequation}) and
(\ref{general-green-boundarycondition}) can be rewritten as
\begin{subeqnarray}
&& \left\{ \frac{\partial}{\partial s} - \frac{\partial^2}{\partial
r^2} + V_{(\kappa,n)}^{\rm free} (r) \right\}
\tilde{\Delta}_{\kappa/2}^{\rm free}
(r,r';s)= 0, \; ({\rm for} \; s>0) \\
&& s \rightarrow 0+: \; \tilde{\Delta}_{\kappa/2}^{\rm free}
(r,r';s) \longrightarrow \delta (r-r'),
\end{subeqnarray}
where the centrifugal potential is
\begin{equation} \label{generalfreepotential}
V_{(\kappa,n)}^{\rm free} (r) = \left\{ \kappa(\kappa+n-2) +
\frac{(n-1)(n-3)}{4} \right\} \frac{1}{r^2} \equiv \frac{g^2}{r^2}.
\end{equation}

To obtain the explicit form of $\tilde{\Delta}_{\kappa/2}^{\rm
free}$, one can resort to a variety of methods (developed to find
the Green function of the one-dimensional Schr\"{o}dinger equation
especially). A particularly elegant method is the one utilizing
quantum canonical transformations, as detailed in Ref.
\cite{leyvraz, tsaur}. As it turns out, for ${\tilde
\Delta}_{\kappa/2}^{\rm free}$, we have a simple closed-form
expression involving the modified Bessel function:
\begin{equation} \label{freeradialgreen}
\tilde{\Delta}_{\kappa/2}^{\rm free} (r,r';s) =
\frac{\sqrt{rr'}}{2s} e^{-\frac{1}{4s} ( r^2 + r'^2 )} I_{\kappa +
\frac{n}{2} -1} \left( \frac{rr'}{2s} \right).
\end{equation}
Since $I_\nu (z) \sim \frac{e^z}{\sqrt{2\pi z}}
[1+O(\frac{1}{|z|})]$ for large $|z|$, the $s\rightarrow 0+$ limit
of this expression is
\begin{equation}
s \rightarrow 0+: \; \tilde{\Delta}_{\kappa/2}^{\rm free} (r,r';s)
\longrightarrow \frac{1}{\sqrt{4\pi s}} e^{-\frac{1}{4s}(r-r')^2} \{
1+O(s) \}.
\end{equation}

The large-$\kappa$ limiting form of $\tilde{\Delta}_{\kappa/2}^{\rm
free} (r,r';s)$ is of interest. Then, due to the large centrifugal
potential term in (\ref{generalfreepotential}), we expect that the
function $\tilde{\Delta}_{\kappa/2}^{\rm free} (r,r';s)$ be
significant (i.e., acquire not-too-small amplitude) only when $s$
lies in the range $0<s \lesssim \frac{A}{\kappa^2}$, $A$ denoting a
constant of $O(1)$. Now, for some large given value of $\kappa$,
suppose that we wish to obtain a systematic approximation of
${\tilde \Delta}_{\kappa/2}^{\rm free} (r,r';s)$ which can be used
for any $s$ satisfying the condition $0<s\kappa^2 \lesssim O(1)$
(this incidentally implies that $s\kappa \ll 1$). Then, to study the
expression in (\ref{freeradialgreen}), we use the known large-order
asymptotic expansion of the modified Bessel function
\cite{abramowitz}
\begin{eqnarray}
\nu \; {\rm large}: \; I_\nu (z) \sim \frac{1}{\sqrt{2\pi\nu}}
\frac{e^{\nu ( \sqrt{1+(z/\nu)^2} + \ln
\frac{z/\nu}{\sqrt{1+(z/\nu)^2}+1} ) }}{ \{ 1+(z/\nu)^2 \}^{1/4}}
\left[ 1 + \frac{3x-5x^3}{24\nu}+ O \left( \frac{1}{\nu^2} \right)
\right], && \nonumber \\
\left( x \equiv \frac{1}{\sqrt{1+(z/\nu)^2}} \in (0,1] \right) &&
\label{bessel-uniformasymptotic}
\end{eqnarray}
with $z=\frac{rr'}{2s}$ and $\nu=\kappa+\frac{n}{2}-1$. [Note that
the expansion (\ref{bessel-uniformasymptotic}) holds {\it uniformly}
with respect to $z$ (i.e., for any small or large $z$), and in the
limit $|z| \rightarrow \infty$ (for fixed $\nu$) goes back to the
asymptotic form given earlier]. Since we are interested in the case
$0<s\kappa^2 \lesssim O(1)$, we may further take the limit
$|\frac{z}{\nu}| = |\frac{rr'}{2s(k+\frac{n}{2}-1)}| \rightarrow
\infty$ with the the formula (\ref{bessel-uniformasymptotic}) (i.e.,
consider an expansion in powers of $|\frac{\nu}{z}|$) and then use
it in (\ref{freeradialgreen}). After some straightforward
calculations, we then obtain the large-$\kappa$ expansion of the
form
\begin{eqnarray}
\tilde{\Delta}_{\kappa/2}^{\rm free} (r,r';s) &\sim&
\frac{1}{\sqrt{4\pi s}} e^{-\frac{1}{4s}(r-r')^2 -
\frac{(\kappa+\frac{n}{2}-1)^2}{rr'} s}
\left\{ 1 + \frac{1}{4rr'}s \right. \nonumber \\
&& \qquad \left. -\frac{(\kappa+\frac{n}{2}-1)^2}{(rr')^2}s^2
+\frac{1}{3} \frac{(\kappa+\frac{n}{2}-1)^4}{(rr')^3}s^3 +
O(\kappa^{-3}) \right\}.  \label{freegreen-uniform}
\end{eqnarray}

We remark that the form (\ref{freegreen-uniform}) may be used to
evaluate a certain quantity which involves, say, the integration of
$\tilde{\Delta}_{\kappa/2}^{\rm free} (r,r';s)$ over the full
$s$-range (i.e., over $s \in (0,\infty)$), as long as $\kappa$ is
constrained to be large. This is because, when $\kappa$ is large,
(i) $\tilde{\Delta}_{\kappa/2}^{\rm free} (r,r';s)$ becomes very
small unless $s\kappa^2 \lesssim O(1)$ (this is also manifest in our
form (\ref{freegreen-uniform})) and (ii) for $s$ satisfying the
condition $0<s\kappa^2 \lesssim O(1)$ we can exploit the expansion
of the form (\ref{freegreen-uniform}) for
$\tilde{\Delta}_{\kappa/2}^{\rm free} (r,r';s)$. In view of this,
the same purpose can be served by rewriting
(\ref{freegreen-uniform}) as
\begin{equation} \label{freegreen-uniform2}
\tilde{\Delta}_{\kappa/2}^{\rm free} (r,r';s) \sim
\frac{1}{\sqrt{4\pi s}} e^{-\frac{1}{4s}(r-r')^2 - \frac{g^2}{rr'}
s} \left\{ 1 -\frac{g^2}{(rr')^2}s^2 + \frac{1}{3}
\frac{g^4}{(rr')^3}s^3 + \cdots \right\},
\end{equation}
where $g^2 \equiv \kappa(\kappa+n-2) + \frac{(n-1)(n-3)}{4} =
(\kappa +\frac{n}{2} -1)^2 - \frac{1}{4}$ (see
(\ref{generalfreepotential})). Note that this series is fully
consistent with our formula (\ref{green-largel}) if we set $V(r)$ to
be equal to $\frac{g^2}{r^2}$.

\section{The $\frac{1}{l}$-expansion with a matrix valued potential}
\label{appmatrixdewitt}

The $\frac{1}{l}$-expansion of $\tilde{\Delta}(r,r;s)$ given in
(\ref{green-largel}) is valid when the potential $V(r)$ is not a
matrix-type. In this Appendix we shall find a more general form which
can be used when $V(r)$ and hence also the Green function
$\tilde{\Delta}(r,r';s)$ are matrix-valued. The coefficient
matrices $b_k (r,r';t)$ in the $\frac{1}{l}$-expansion will now have
to satisfy the matrix equations in (\ref{recursion1}) and
(\ref{recursion2}). Choosing a new independent variable $u$ (instead
of $r$) by setting $r=r'+tu$ and writing $b_k (r'+tu,r';t) \equiv
\bar{b}_k (u,r';t)$, we may recast these equations as
\begin{subeqnarray}
O(l^2) &:& \partial_t \bar{b}_0 (u,r';t) + U(r'+tu) \bar{b}_0
(u,r';t) = 0, \slabel{mrecursion1} \\
O(l^{2-2k}) &:& \partial_t \bar{b}_k (u,r';t) + U(r'+tu) \bar{b}_k
(u,r';t) - \frac{1}{t^2} \partial_\mu^2 \bar{b}_{k-1} (u,r';t) = 0,
\slabel{mrecursion2} \\
&& \qquad\qquad\qquad\qquad\qquad\qquad\qquad\qquad
(k=1,2,3,\cdots). \nonumber
\end{subeqnarray}
The solution of (\ref{mrecursion1}), satisfying the boundary
condition $\bar{b}_0 (u,r';t=0) = 1$, is
\begin{equation} \label{mb0}
\bar{b}_0 (u,r';t) = P \left[ e^{-\int_0^t U(r'+t_1 u) dt_1}
\right],
\end{equation}
where $P[\cdots]$ denotes the $t$-ordering. Setting $u=0$ in
(\ref{mb0}) then gives
\begin{equation} \label{mb0c}
b_0 (r,r;t) = e^{-t U(r)},
\end{equation}
i.e, our formula (\ref{green-eqn-boundary}) for the coincidence
limit of $b_0 (r,r';t)$ holds even when $U(r)$ is matrix valued.

To solve (\ref{mrecursion2}) for higher-order coefficient $\bar{b}_k
(u,r';t)$ $(k=1,2,\cdots)$, one may follow the steps similar to
(\ref{newset}) and (\ref{recursion4}) --- rewrite the equations as
those for the matrix functions $\tilde{b}_k (u,r';t)$ which are
obtained through multiplying $\bar{b}_k (u,r';t)$ from the left by
the inverse of the $t$-ordered exponential matrix in (\ref{mb0}).
But what we need here is only the coincidence limits, i.e., $b_k
(r,r;t) \equiv \bar{b}_k (u=0,r'=r;t)$ for small $k$, and for the
latter it is actually simpler to obtain the desired expressions by
considering the $u=0$ limits of our differential equations
(\ref{mrecursion1}) and (\ref{mrecursion2}) and of their derivative
relations \cite{lee2}. Specifically, for $\bar{b}_1 (u=0,r,t)$, we
have the equation (by setting $u=0$ with (\ref{mrecursion2}))
\begin{equation} \label{mb1ceq}
\partial_t \bar{b}_1 (0,r;t) + U(r) \bar{b}_1 (0,r;t) -
\frac{1}{t^2} \left. \left[ \partial_u^2 \bar{b}_0 (u,r;t) \right]
\right|_{u=0} = 0,
\end{equation}
which can readily be integrated (with the `initial' condition
$\bar{b}_1 (u,r;t=0) = 0$) only if the expression for $\left. \left[
\partial_u^2 \bar{b}_0 (u,r;t) \right] \right|_{u=0}$ is known.
Then, setting $u=0$ in the relations obtained after single and twice
differentiations of (\ref{mrecursion1})) with respect to $u$, we
have
\begin{eqnarray}
&& \partial_t \left. \left[ \partial_u \bar{b}_0 (u,r;t) \right]
\right|_{u=0} + U(r) \left. \left[ \partial_u \bar{b}_0 (u,r;t)
\right] \right|_{u=0} + tU'(r) e^{-tU(r)} = 0, \label{dmb0} \\
&& \partial_t \left. \left[ \partial_u^2 \bar{b}_0 (u,r;t) \right]
\right|_{u=0} + U(r) \left. \left[ \partial_u^2 \bar{b}_0 (u,r;t)
\right] \right|_{u=0} \nonumber \\
&& \qquad\qquad\qquad + 2tU'(r) \left. \left[ \partial_u \bar{b}_0
(u,r;t) \right] \right|_{u=0} + t^2 U''(r) e^{-tU(r)} = 0,
\label{ddmb0}
\end{eqnarray}
where the result in (\ref{mb0c}) has been used. From (\ref{dmb0}) it
follows that
\begin{equation}
\left. \left[ \partial_u \bar{b}_0 (u,r;t) \right] \right|_{u=0} =
e^{-tU(r)} \int_0^t dt_1 \; t_1 e^{t_1 U(r)} U'(r) e^{-t_1 U(r)}.
\end{equation}
Using this result, we can go on to integrate (\ref{ddmb0}) to obtain
\begin{eqnarray}
\left. \left[ \partial_u^2 \bar{b}_0 (u,r;t) \right] \right|_{u=0}
&=& e^{-tU(r)} \left\{ 2\int_0^t dt_1 \; t_1 e^{t_1 U(r)} U'(r)
e^{-t_1 U(r)} \int_0^{t_1} dt_2 \; t_2 e^{t_2 U(r)} U'(r) e^{-t_2
U(r)} \right. \nonumber \\
&& \qquad\qquad \left. - \int_0^t dt_1 \; t_1^2 e^{t_1 U(r)} U''(r)
e^{-t_1 U(r)} \right\}.
\end{eqnarray}
Now, by using this result in (\ref{mb1ceq}) and integrating the
resulting equation, we find the expression for the coincidence
limit $b_1 (r,r;t)(=\bar{b}_1 (0,r;t))$:
\begin{eqnarray}
b_1 (r,r;t) &=& e^{-tU(r)} \int_0^t dt_1 \frac{1}{t_1^2} \left\{
2\int_0^{t_1} dt_2 \; t_2 e^{t_2 U(r)} U'(r) e^{-t_2 U(r)}
\int_0^{t_2} dt_3 \; t_3 e^{t_3 U(r)} U'(r) e^{-t_3 U(r)} \right.
\nonumber \\
&& \qquad\qquad\qquad \left. - \int_0^{t_1} dt_2 \; t_2^2 e^{t_2
U(r)} U''(r) e^{-t_2 U(r)} \right\}. \label{mb1c}
\end{eqnarray}

The desired $\frac{1}{l}$-expansion, which generalizes
(\ref{green-largel}) to the case of a matrix-valued potential,
follows upon using the results (\ref{mb0c}) and (\ref{mb1c}) with
(\ref{ansatz}). It has the following structure:
\begin{eqnarray}
&& \tilde{\Delta} (r,r;s) = \frac{1}{\sqrt{4\pi s}} e^{-s V(r)}
\left[ 1+ \int_0^s ds_1 \frac{1}{s_1^2} \left\{ 2\int_0^{s_1} ds_2
\; s_2\, e^{s_2 V(r)} V'(r) e^{-s_2 V(r)} \right. \right. \nonumber \\
&& \left. \times \int_0^{s_2} ds_3 \; s_3\, e^{s_3 V(r)} V'(r) e^{-s_3
V(r)} - \int_0^{s_1} ds_2 \; s_2^2\, e^{s_2 V(r)} V''(r) e^{-s_2 V(r)}
\right\} + O \left( \frac{1}{l^4} \right) \Bigg].
\end{eqnarray}

\section{Use of the Euler-Maclaurin Summation Formula} \label{appeuler}

In this appendix 
we explain how the Euler-Maclaurin summation formula
\cite{abramowitz, kahn} can be used to sum the various partial-wave
contributions to the effective action. First, we present the related
mathematical theory. Let $f(x)$ be a function with continuous
derivatives up to order $2m+2$ for $x \in [a,b]$, where $a$ and $b$
are integers. Then, for the sum
\begin{equation} \label{sumdefinition}
\sum_{n=a}^b f(n) \equiv f(a) + f(a+1) + \cdots + f(b-1) + f(b),
\end{equation}
we have the Euler-Maclaurin summation formula
\begin{equation}
\sum_{n=a}^b f(n) = \int_a^b f(x) dx + \frac{1}{2} [f(a)+f(b)] +
\sum_{k=1}^m \frac{B_{2k}}{(2k)!} [f^{(2k-1)}(b) - f^{(2k-1)}(a)] +
R_m ,
\end{equation}
where $B_j$ are the Bernoulli numbers $(B_2=\frac{1}{6}, \;
B_4=-\frac{1}{30}, \; B_6=\frac{1}{42}, \cdots)$ and the remainder
term is
\begin{equation}
R_m = \frac{(b-a)B_{2m+2}}{(2m+2)!} f^{(2m+2)}(\bar{\theta}), \qquad
\textrm{for some } \bar{\theta} \in (a,b).
\end{equation}
This formula will be particularly useful to evaluate the sum of
slowly varying terms with decreasing derivatives. (A nice treatment
on this formula, including the derivation, is given in Ref.
\cite{kahn}).

As for the partial wave sums in our work, we may use
(\ref{sumdefinition}) in the form (here $l=\frac{n}{2}$, $n$ being
integers)
\begin{eqnarray}
\sum_{l=L+\frac{1}{2}}^\infty f(l) &=& \sum_{n=2L}^\infty f \left(
\frac{n}{2} \right) - f(L) \nonumber \\
&=& \left\{ \int_{2L}^\infty f\left( \frac{x}{2} \right) dx +
\frac{1}{2} f(L) - \frac{1}{12} \left. \left[ \frac{d}{dx} f \left(
\frac{x}{2} \right) \right] \right|_{x=2L} + \cdots \right\} - f(L)
\nonumber \\
&=& 2\int_L^\infty f(l) dl - \frac{1}{2} f(L) - \frac{1}{24} f'(L) +
\cdots, \label{euler-maclaurina}
\end{eqnarray}
assuming $f(\infty)=f'(\infty)=0$, etc. To deal with various
$l$-sums appearing in (\ref{igcase1explicit}), we may here take
\begin{equation} \label{sumform}
f(l) = e^{-s (a_2 l^2 +a_1 l + a_0 )} (b_0 + b_1 l + b_2 l^2 +
\cdots ), \qquad (a_2>0).
\end{equation}
Then, to perform the $l$-integral $\int_L^\infty f(l)dl$, we change
the integration variable from $l$ to $t$ by setting
\begin{equation}
l = \frac{t}{\sqrt{2a_2s}} - \frac{a_1}{2a_2}.
\end{equation}
This will put the function (\ref{sumform}) in a simpler form, i.e.,
\begin{equation}
f(l) \longrightarrow \tilde{f}(t) = e^{-t^2} (\tilde{b}_0 +
\tilde{b}_1 t + \tilde{b}_2 t^2 + \cdots )
\end{equation}
and the resulting integrals can be done by using the formula
\cite{abramowitz}
\begin{equation}
\int_T^\infty dt e^{-t^2} t^n =
\begin{cases}{ \displaystyle \frac{1}{2} \left( \frac{n-1}{2} \right)! e^{-T^2}
\sum_{k=0}^{\frac{n-1}{2}} \frac{T^{2k}}{k!}, \qquad (n= \textrm{odd
integer}) \cr \displaystyle \frac{1}{2} \left[ \left( \frac{n-1}{2}
\right)! \, {\rm erfc}(T) + T e^{-T^2} \sum_{k=0}^{\frac{n}{2}-1}
\left( k+ \frac{3}{2} \right)_{-k+\frac{n}{2}-1} T^{2k} \right], \cr
\qquad\qquad\qquad\qquad\qquad\qquad\quad (n= \textrm{even integer})
} \end{cases}
\end{equation}
where $(a)_n \equiv a(a+1)...(a+n-1)$ is the Pochhammer symbol. Now
the $l$-sums in (\ref{igcase1explicit}) can be performed explicitly.
If we discard contributions vanishing for sufficiently large $L$,
these sums for the first two terms in the right hand side of
(\ref{igcase1explicit}) read
\begin{eqnarray}
&& \sum_{l=L+\frac{1}{2}}^\infty (2l+1)(2l+2)
\tilde{\Delta}_{(l,j=l+\frac{1}{2})} (r,r;s) \nonumber \\
&& \qquad = e^{s(\frac{1}{4r^2} + 2f(r) - 2r^2f(r)^2)} {\rm erfc}
\left( \frac{\sqrt{s}}{r} (2L+1+r^2f(r)) \right) \left\{
\frac{r^3}{8s^2} + \frac{r}{32s} (-1 - 8r^2f(r) \right. \nonumber \\
&& \qquad\qquad\qquad \left. + 8r^4f(r)^2) + \frac{1}{16} (
f(r)(r-4r^4f'(r)) - 5r^3f(r)^2 - r^5f'(r)^2) \right\} \nonumber \\
&& \quad\qquad + \frac{1}{\sqrt{4\pi s}} e^{-s( \frac{4L(L+1) +
\frac{3}{4}}{r^2} + 4Lf(r) + 3r^2f(r)^2 )} \left\{ \frac{r^2L}{s} +
\left( -2L^2 + \frac{r^2}{s} - \frac{r^4}{2s}f(r) \right) + \left(
-\frac{43L}{12} \right. \right. \nonumber \\
&& \qquad\qquad\qquad \left. + \frac{2sL^3}{3r^2} +
\frac{16s^2L^5}{3r^4} \right) + \left( -\frac{37}{24}
-\frac{r^2}{8}f(r) - \frac{2r^3}{3}f'(r) - \frac{r^4}{6}f''(r) +
\frac{sL^2}{r^2} \right. \nonumber \\
&& \qquad\qquad\qquad  \left. - \frac{sL^2}{3}f(r) -
\frac{8rsL^2}{3}f'(r) - \frac{2r^2sL^2}{3}f''(r) +
\frac{24s^2L^4}{r^4} - \frac{8s^2L^4}{3r^2}f(r) \right. \nonumber
\\
&& \qquad\qquad\qquad \left. - \frac{16s^2L^4}{3r}f'(r) -
\frac{32s^3L^6}{3r^6} \right) + \left( \frac{sL}{2r^2}
-\frac{sL}{6}f(r) -3r^2sLf(r)^2 -\frac{r^4sL}{2}f'(r)^2 \right.
\nonumber \\
&& \qquad\qquad\qquad - \frac{8rsL}{3}f'(r) -
\frac{14r^3sL}{3}f(r)f'(r) - \frac{2r^2sL}{3}f''(r)
-\frac{2r^4sL}{3}f(r)f''(r) \nonumber \\
&& \qquad\qquad\qquad + \frac{122s^2L^3}{3r^4} -
\frac{20s^2L^3}{3}f(r)^2 + \frac{4r^2s^2L^3}{3}f'(r)^2 -
\frac{20s^2L^3}{3r^2}f(r) - \frac{32s^2L^3}{3r}f'(r) \nonumber \\
&& \qquad\qquad\qquad \left. \left. - \frac{16rs^2L^3}{3}f(r)f'(r) +
\frac{4s^2L^3}{3}f''(r) -\frac{48s^3L^5}{r^6} +
\frac{32s^3L^5}{3r^3}f'(r) + \frac{64s^4L^7}{9r^8} \right) \right\}
\nonumber \\
&& \qquad \equiv {\cal F}_1 (r,s), \label{suma} \\\nonumber\\
&& \sum_{l=L+\frac{1}{2}}^\infty (2l+1)(2l+2)
\tilde{\Delta}_{(l+\frac{1}{2},j=l)} (r,r;s) \nonumber \\
&& \qquad = e^{s(\frac{1}{4r^2} + 2f(r) - 2r^2f(r)^2)} {\rm erfc}
\left( \frac{\sqrt{s}}{r} (2L+2-r^2f(r)) \right) \left\{
\frac{r^3}{8s^2} + \frac{r}{32s} (-1 - 8r^2f(r) \right. \nonumber \\
&& \qquad\qquad\qquad \left. + 8r^4f(r)^2) + \frac{1}{16} (
f(r)(r-4r^4f'(r)) -
5r^3f(r)^2 - r^5f'(r)^2) \right\} \nonumber \\
&& \quad\qquad + \frac{1}{\sqrt{4\pi s}} e^{-s( \frac{4L(L+2) +
\frac{15}{4}}{r^2} - (4L+6)f(r) + 3r^2f(r)^2 )} \left\{
\frac{r^2L}{s} + \left( -2L^2 + \frac{r^2}{2s} - \frac{r^4}{2s}f(r)
\right) \right. \nonumber \\
&& \qquad\qquad\qquad + \left( -\frac{43L}{12} + \frac{2sL^3}{3r^2}
+ \frac{16s^2L^5}{3r^4} \right) + \left( -\frac{4}{3}
+\frac{r^2}{8}f(r) + \frac{2r^3}{3}f'(r) + \frac{r^4}{6}f''(r)
\right. \nonumber \\
&& \qquad\qquad\qquad  + \frac{2sL^2}{r^2} + \frac{sL^2}{3}f(r) +
\frac{8rsL^2}{3}f'(r) + \frac{2r^2sL^2}{3}f''(r) +
\frac{32s^2L^4}{r^4} + \frac{8s^2L^4}{3r^2}f(r) \nonumber \\
&& \qquad\qquad\qquad \left. + \frac{16s^2L^4}{3r}f'(r) -
\frac{32s^3L^6}{3r^6} \right) + \left( \frac{2sL}{r^2}
+\frac{5sL}{6}f(r) -3r^2sLf(r)^2 -\frac{r^4sL}{2}f'(r)^2 \right.
\nonumber \\
&& \qquad\qquad\qquad + \frac{16rsL}{3}f'(r) -
\frac{14r^3sL}{3}f(r)f'(r) - \frac{4r^2sL}{3}f''(r)
-\frac{2r^4sL}{3}f(r)f''(r) \nonumber \\
&& \qquad\qquad\qquad + \frac{218s^2L^3}{3r^4} -
\frac{20s^2L^3}{3}f(r)^2 + \frac{4r^2s^2L^3}{3}f'(r)^2 +
\frac{28s^2L^3}{3r^2}f(r) + \frac{64s^2L^3}{3r}f'(r) \nonumber \\
&& \qquad\qquad\qquad \left. \left. - \frac{16rs^2L^3}{3}f(r)f'(r) -
\frac{4s^2L^3}{3}f''(r) -\frac{208s^3L^5}{3r^6} -
\frac{32s^3L^5}{3r^3}f'(r) + \frac{64s^4L^7}{9r^8} \right) \right\}
\nonumber \\
&& \qquad \equiv {\cal F}_2 (r,s). \label{sumb}
\end{eqnarray}
If we set $f(r)=0$ in these expression, they represent the $l$-sums
for the last two terms of (\ref{igcase1explicit}) (i.e., for those
containing the factors $e^{-s V_{(l)}^0 (r)}$ and $e^{-s
V_{(l+\frac{1}{2})}^0 (r)}$). In this way we obtain the explicit
double-integral representation for $\Gamma_{J>J_L}^\epsilon (A;m)$
of the form
\begin{equation} \label{appeulerresult}
\Gamma_{J>J_L}^\epsilon (A;m) = -\int_0^\infty dr \int_0^\infty ds
\frac{e^{-m^2s}}{s} s^\epsilon \left[ {\cal F}_1(r,s) + {\cal
F}_2(r,s) - {\cal F}_1(r,s)|_{f(r)=0} - {\cal F}_2(r,s)|_{f(r)=0}
\right].
\end{equation}
The $l$-sums in (\ref{igcase2}) can be performed in a similar
manner.

\end{document}